\documentclass[journal,onecolumn,letterpaper]{IEEEtran}
\usepackage{amsthm,mathtools,xcolor}
\usepackage{times,newtxtext,newtxmath}
\usepackage{booktabs,array,enumitem,microtype}
\usepackage[hidelinks]{hyperref}
\usepackage{cite}

\newcommand{\Id}{\mathbb{I}}
\newcommand{\Tr}{\operatorname{Tr}}

\newcommand{\E}{\mathbb{E}}

\newtheorem{theorem}{Theorem}[section]
\newtheorem{lemma}[theorem]{Lemma}
\newtheorem{proposition}[theorem]{Proposition}
\newtheorem{corollary}[theorem]{Corollary}
\newtheorem{remark}[theorem]{Remark}



\title{Adaptive Local Measurements For Quantum Learning Can Exponentially Outperform Nonadaptive Global Measurements}
\author{
Alireza Goldar\IEEEauthorrefmark{1}, Zhen Qin\IEEEauthorrefmark{2}, Zhihui Zhu\IEEEauthorrefmark{3}, Zhe-Xuan Gong\IEEEauthorrefmark{4}, and Michael B. Wakin\IEEEauthorrefmark{1}
\thanks{\IEEEauthorrefmark{1} Department of Electrical Engineering, Colorado School of Mines, Golden, Colorado 80401, USA}
\thanks{\IEEEauthorrefmark{2} Michigan Institute for Computational Discovery and Engineering, Department of Electrical Engineering and Computer Science, and Department of Statistics, University of Michigan, Ann Arbor, Michigan 48109, USA}
\thanks{\IEEEauthorrefmark{3} Department of Computer Science and Engineering, The Ohio State University, Columbus, Ohio 43201, USA}
\thanks{\IEEEauthorrefmark{4} Department of Physics, Colorado School of Mines, Golden, Colorado 80401, USA}
\thanks{This material is based upon work supported by the National Science Foundation under Awards CCF-2106834, CCF-2241298, and ECCS-2409701.}
}

\begin{document}
\maketitle

\begin{abstract}
Can measurements chosen adaptively, in response to earlier outcomes, be exponentially more informative than measurements fixed in advance? In many quantum settings, adaptivity helps at most polynomially; known exponential advantages require coherent measurements across copies or restricted measurements. We show an exponential advantage measuring one copy at a time, with no quantum memory kept between state copies, even when the nonadaptive learner may perform arbitrary global measurements on each copy. We construct explicit families of multi-qubit states in which a scalar parameter is encoded in one of exponentially many candidate observables, with an auxiliary marker indicating which one. An adaptive protocol identifies the relevant observable and then measures it using only single-qubit measurements and one round of adaptation. At fixed accuracy, it needs logarithmically many copies as the qubit count grows, whereas every nonadaptive protocol needs exponentially many. The same mechanism yields an exponential gap in local asymptotic minimax mean squared error constants, tight up to the exponential base. The states are separable and classical in label-dependent product bases; no entanglement or other nonclassical correlations are required. The gap comes from measurement incompatibility: the candidate observables cannot be measured jointly, so a fixed measurement must divide its information budget among them, while adaptivity selects the informative one. Accordingly, the separation disappears when the candidate observables commute, and cannot occur with fully observed classical samples. Measurement incompatibility thus emerges as a resource for quantum learning, distinct from entanglement and quantum memory, that classical feedback alone can exploit.
\end{abstract}


\section{Introduction}

\subsection{Overview}

Quantum state tomography (QST) seeks to reconstruct an unknown quantum state from measurements performed on identically prepared copies of that state. While QST provides a complete characterization of the system, the cost of reconstructing a generic many-qubit state grows exponentially with system size, motivating approaches that exploit structural assumptions, more informative measurements, or narrower learning objectives, such as predicting observables without reconstructing the full state~\cite{AnshuArunachalam2024Survey,HaahHarrowJiWuYu2017,ODonnellWright2016}. One particularly appealing possibility is to choose measurements adaptively: outcomes observed on early copies are used to decide which measurements should be performed on later copies. Adaptive measurement selection has been studied from Bayesian, experimental, and active learning perspectives and can provide substantial practical or asymptotic improvements in some regimes~\cite{HuszarHoulsby2012,MahlerRozemaDarabiFerrieBlumeKohoutSteinberg2013,GranadeFerrieFlammia2017,LangeKebricBuserSchollwockGrusdtBohrdt2023,AasenGarttner2026}. The central question, however, is not simply whether adaptivity can help, but under what combination of state structure, measurement resources, and loss function it can produce a fundamental statistical advantage.

Recent theory makes this dependence particularly clear. For tomography of an arbitrary mixed state using general single copy measurements (i.e., measuring one copy of the state at a time), Chen et al.~\cite{ChenHuangLiLiuSellke2023} showed that adaptivity does not improve the optimal sample complexity scaling in trace distance, although it does improve the scaling for infidelity. Lowe and Nayak~\cite{LoweNayak2025} established further lower bounds for single copy tomography, including adaptive protocols whose measurements are drawn from sufficiently small fixed families. At the same time, adaptivity can improve asymptotic infidelity scaling for qubit tomography~\cite{MahlerRozemaDarabiFerrieBlumeKohoutSteinberg2013} and can yield an exponential advantage for structured state families under restricted measurements~\cite{GoldarQinZhuGongWakin2026}. Thus whether adaptivity helps must be studied together with the state family, the measurement model, and the loss function.

Very recent work has sharpened this question from both sides. An exponential adaptive/nonadaptive separation is now known for Pauli shadow tomography with \emph{two copy} measurements~\cite{ChenGongZhang2024}, while within the single copy model the known adaptive advantages are polynomial for moment estimation~\cite{Liu2026Moments} and for stabilizer learning~\cite{BittelEisertGongMeleSchatzki2026}; the latter work notes that provable separations from nonadaptive strategies remain rare and asks for structural conditions under which adaptivity provably helps. The present paper gives such a condition together with an exponential single copy separation.

The measurement model is especially important here. For example, Pauli measurements are experimentally convenient and play a central role in tomography, but they do not represent the full power of single copy measurements. Recent work has established a separation between Pauli measurements and more general single copy measurements for worst case tomography~\cite{AcharyaDharmavarapuLiuYu2025}, while related work on quantum state testing shows that restricting the admissible measurement family can substantially change statistical complexity~\cite{LiuAcharya2024}. Consequently, an adaptive advantage established under a Pauli measurement restriction need not represent an intrinsic advantage over all nonadaptive single copy protocols: a more general fixed measurement may remove the bottleneck that made adaptation useful.

In previous work~\cite{GoldarQinZhuGongWakin2026}, we constructed a structured prefix/tree family for which, under local product Pauli measurements, an adaptive protocol has polynomial sample complexity while every nonadaptive Pauli design requires exponentially many copies in the worst case. That result demonstrated that sequential basis selection alone can fundamentally change sample complexity scaling within a suitable structured family, but it kept the measurement class fixed to local product Pauli measurements. It therefore left open a stronger question: \emph{can an exponential adaptive advantage survive when the nonadaptive learner is allowed measurements substantially more general than product Pauli settings?} The present paper addresses this question using new state families designed to make the underlying information allocation mechanism analytically transparent.

The basic mechanism is simple. Each state contains a marker register and a signal register. A hidden label can be inferred from the marker, and this label determines which signal observable is informative about an unknown scalar parameter. An adaptive protocol can therefore use early copies to identify the hidden label and then concentrate subsequent measurements on the corresponding signal observable. In contrast, any nonadaptive protocol must choose its measurements before knowing the label and must instead distribute its measurement effort across many candidate signal observables. Locally around the reference point used in our analysis, the signal state itself is the same for all hidden labels; what changes with the label is the tangent direction in which the state varies with the scalar parameter. This makes Fisher information a natural tool for quantifying how well a fixed measurement can cover the possible local tangent directions simultaneously.

We first study a binary hidden label construction. Under product Pauli measurements, it exhibits an exponential separation between adaptive and nonadaptive protocols. We then enlarge the nonadaptive measurement class to arbitrary product positive operator valued measures (POVMs) on the individual qubits and show that, locally around the maximally mixed signal state, the exponential separation persists. Thus the separation is not simply an artifact of restricting the measurements to Pauli bases. However, the binary construction does \emph{not} survive a further enlargement to arbitrary global single copy POVMs. Its support dependent signal observables commute and therefore admit a common eigenbasis. A fixed global measurement in this basis can retain the statistics associated with all candidate hidden labels simultaneously; after the marker identifies the relevant label, the corresponding statistic can be selected in classical post processing. In this case, adaptive retargeting of the signal measurement is unnecessary.

This failure identifies what the binary construction is missing. At the global level, hiding the informative observable is not sufficient if all candidate observables can be accessed through one common measurement. We therefore introduce a ternary construction in which the hidden label selects among three Pauli observables at each signal site. Because candidate signal observables no longer share a common eigenbasis, no single fixed measurement can simultaneously access all candidate signal statistics. The absence of a common eigenbasis is only a qualitative distinction; the nonadaptive lower bound requires a quantitative information constraint. That constraint is provided by an aggregate Fisher information bound showing that any fixed global single copy POVM has a limited total information budget across the candidate signal observables. Consequently, every fixed nonadaptive protocol has at least one hidden label for which the available Fisher information is exponentially small, whereas an adaptive protocol can first identify that label from the marker and then measure the matched signal observable. If \(n\) denotes the total number of qubits in each copy, we obtain a \(2^{\Omega(n)}\) lower bound on the ratio between the nonadaptive and adaptive leading constants in the local asymptotic minimax mean squared error (MSE). The local asymptotic minimax MSE criterion is taken over a shrinking neighborhood of the reference value of the scalar parameter and is defined precisely in Section~\ref{sec:framework}. The same information budget also yields a finite sample separation at fixed accuracy. The adaptive protocol needs only \(O(\log n)\) state copies, using single qubit Pauli measurements and one round of adaptation, whereas every nonadaptive protocol using arbitrary global single copy POVMs requires \(2^{\Omega(n)}\) copies (Theorem~\ref{thm:ternary_finite_sample}). This is the precise sense of the title: adaptive local measurements exponentially outperform nonadaptive global ones.

An important feature of the construction is its simplicity. The state families introduced in this paper are fully separable. Moreover, when each marker qubit is grouped with its corresponding signal qubit, the ternary family admits an exact matrix product operator (MPO) representation~\cite{VerstraeteGarciaRipollCirac2004,QinJamesonGongWakinZhu2024} with constant bond dimension. Thus the exponential separation occurs within a compactly representable state family and is driven by hidden measurement structure and adaptive information allocation. In particular, the adaptive separation does not require any of the following: entanglement or any nonclassical correlation in the unknown state,  collective measurements across copies, or a computationally expensive adaptive rule.

Since the states are separable, it is natural to ask what, if anything, is quantum about the separation. Section~\ref{sec:full_separability} answers this in three steps. First, every state in both families is diagonal in a label dependent local product basis. Thus each state is not only fully separable but also classical, with zero quantum discord. Second, the global separation has no classical analogue: in any classical statistical model in which a nonadaptive learner observes each sample completely, that learner can simulate every adaptive protocol in post processing. Third, what remains is measurement incompatibility. The hidden label determines which signal measurement is informative, but a nonadaptive learner must choose its measurement before that label is known. An adaptive learner, by contrast, can first learn the label and then choose the appropriate signal measurement. This is exactly the distinction studied in quantum state discrimination between having the label before the measurement and receiving it only afterward. An advantage gained from knowing the label prior to measurement choice is closely tied to witness measurement incompatibility~\cite{BallesterWehnerWinter2008,GopalWehner2010,CarmeliHeinosaariToigo2018,CarmeliHeinosaariToigo2019}. Quantitatively, the limitation on the nonadaptive learner is captured by our aggregate Fisher information bound, which is an instance of the Gill--Massar inequality at the maximally mixed state~\cite{GillMassar2000}. The resource behind the separation is therefore measurement incompatibility, which is distinct from entanglement, from nonclassical correlations in the state, and from quantum memory across copies.

\subsection{Related work}
\label{sec:related_work}

Adaptive quantum tomography has a substantial history. Husz\'ar and Houlsby~\cite{HuszarHoulsby2012} formulated adaptive Bayesian tomography in which measurement choices are selected to maximize expected information gain. Mahler et al.~\cite{MahlerRozemaDarabiFerrieBlumeKohoutSteinberg2013} demonstrated that a simple adaptive step can improve the asymptotic infidelity scaling for qubit tomography from \(O(N^{-1/2})\) to \(O(N^{-1})\), where \(N\) denotes the number of copies, and Granade et al.~\cite{GranadeFerrieFlammia2017} developed practical adaptive methods for larger estimation problems. More recently, Lange et al.~\cite{LangeKebricBuserSchollwockGrusdtBohrdt2023} developed an active learning tomography scheme that, after an initial set of measurements, adaptively selects subsequent measurement bases to maximize information gain. They empirically demonstrated improved many-body state reconstruction relative to randomly chosen basis configurations using the same measurement budget. Their emphasis is on adaptive reconstruction algorithms, whereas our goal is to establish a theoretical adaptive/nonadaptive separation and identify the information allocation mechanism responsible for it. Theoretical results also show that the benefits of adaptation depend strongly on the error metric and measurement model. For \(d\)-dimensional mixed states under single copy measurements, Chen et al.~\cite{ChenHuangLiLiuSellke2023} showed that achieving trace distance error \(\epsilon\) requires \(\Theta(d^3/\epsilon^2)\) copies even with adaptive measurements. For infidelity \(\gamma\), however, adaptivity improves the dependence on the accuracy parameter: an adaptive protocol achieves \(\widetilde{O}(d^3/\gamma)\) copies, whereas any nonadaptive protocol requires \(\Omega(d^3/\gamma^2)\) copies. Thus the improvement is in the dependence on \(\gamma\), while the \(d^3\) dependence remains unchanged. Aasen and G\"arttner~\cite{AasenGarttner2026} showed that readout noise can remove the asymptotic scaling improvement of adaptive protocols in a pure state infidelity setting. 

Collectively, these results reinforce the importance of specifying the state class, error metric, and measurement resources when assessing the value of adaptivity. Separations between adaptive and nonadaptive single copy protocols are therefore not new in themselves: the qubit result of~\cite{MahlerRozemaDarabiFerrieBlumeKohoutSteinberg2013}, which builds on the two stage scheme of Gill and Massar~\cite{GillMassar2000}, and its extension to all dimensions in~\cite{ChenHuangLiLiuSellke2023} are separations in the dependence on the accuracy parameter (sample complexity \(1/\gamma\) versus \(1/\gamma^2\) in infidelity). Separations in the dependence on system size are more recent and are the relevant comparison here. Fawzi et al.~\cite{FawziFlammarionGarivierOufkir2023} showed that when a \(d\)-dimensional state is promised to be diagonal in one of \(m\) known bases, locating the basis and then measuring in it costs \(O(d\log m/\epsilon^2)\) copies, whereas nonadaptive protocols need \(\Omega(\min\{md/(\epsilon^2\log m),\,d^2/\epsilon^2\})\), a polynomial gap in \(d\) with an adaptive cost that remains linear in \(d\); they also note that no such separation exists in classical testing. Subsequent work established an exponential adaptive/nonadaptive separation for Pauli shadow tomography with two copy measurements~\cite{ChenGongZhang2024}. Polynomial single copy adaptive advantages have also been shown for moment estimation~\cite{Liu2026Moments}, while Bittel et al.~\cite{BittelEisertGongMeleSchatzki2026} developed an optimal adaptive single copy learner for stabilizer states. The present paper reaches a different regime: single copy measurements, an adaptive cost that is logarithmic in the number of qubits, and a nonadaptive cost that is exponential in it.

Our work is also related to the theory of restricted single copy measurements. Optimal tomography with unrestricted collective measurements across copies is well understood at the level of sample complexity~\cite{ODonnellWright2016,HaahHarrowJiWuYu2017}, but such measurements require substantially stronger experimental resources. Within the single copy model, Lowe and Nayak~\cite{LoweNayak2025} derived lower bounds for nonadaptive tomography and for broad classes of adaptive measurement protocols with implementation restrictions. Liu and Acharya~\cite{LiuAcharya2024} developed information theoretic lower bounds for quantum state testing under restricted measurement families. Particularly relevant to the present work, Acharya et al.~\cite{AcharyaDharmavarapuLiuYu2025} showed that Pauli measurements are statistically suboptimal relative to more general single copy measurements for worst case tomography. These results motivate our progression from product Pauli measurements, to arbitrary product POVMs, and finally to arbitrary global measurements within each copy. The Pauli restricted separation of~\cite{GoldarQinZhuGongWakin2026} rests on counting measurement settings; once arbitrary POVMs are allowed there are no settings to count, and Section~\ref{sec:ternary_global_separation} replaces the count by an information budget over candidate observables.

We offer a brief remark here on terminology. The results above concern tomography, which asks for the whole density matrix, so that its sample complexity carries the cost of estimating \(d^2-1\) parameters, and the adaptive advantages known for it improve the dependence on the accuracy rather than on the dimension. The problem studied here is a \emph{quantum parameter estimation} problem in the sense of Helstrom and Holevo~\cite{Helstrom1969,Holevo1982,Paris2009}: a single real parameter is to be estimated from many copies, with mean squared error controlled by Fisher information, and the hidden label is a discrete \emph{nuisance parameter}, an unknown that is not of interest in itself but decides which measurement is informative~\cite{SuzukiYangHayashi2020}. In the vocabulary of quantum learning theory the task is property estimation over a structured family, not tomography or testing. Section~\ref{sec:reconstruction_metrics} relates the scalar accuracy used here to state reconstruction metrics.

A separate but complementary line of work exploits low complexity structure in the unknown state. Compressed sensing approaches use low rank to reduce tomographic complexity~\cite{GrossLiuFlammiaBeckerEisert2010,FlammiaGrossLiuEisert2012}, while tensor network methods exploit matrix product states, matrix product operators, and related representations~\cite{CramerPlenioFlammiaSommaGrossBartlettLandonCardinalPoulinLiu2010,VerstraeteGarciaRipollCirac2004}. Recent work has developed tomography and learning methods for tensor network models using local informationally complete measurements, single qubit measurements, locally purified density operators, randomized measurements, and higher dimensional tensor networks~\cite{JamesonQinGoldarWakinZhuGong2024,KuzminMikhailovaDyakonovStraupe2024,GuoYang2024,VottoLjubotinaLancienCiracZollerSerbynPiroliVermersch2025,QinZhu2026}. The exact constant bond dimension MPO representation of our family shows that an exponential adaptive advantage can arise even within a highly structured and compactly representable class of states.

Fisher information and Cram\'er--Rao methods have long been used to evaluate and optimize quantum measurement designs~\cite{KosutWalmsleyRabitz2004,JamesonQinGoldarWakinZhuGong2024}. In the present work, Fisher information plays a complementary role. Rather than optimizing one measurement for one known local tangent direction, we use an aggregate information bound to show that a single fixed measurement cannot simultaneously provide large information for all candidate signal observables in the ternary family. This information allocation obstruction is the technical mechanism behind our global nonadaptive lower bound.

Two classical facts enter the analysis. The aggregate Fisher bound behind our lower bound is the Gill--Massar inequality~\cite{GillMassar2000} specialized to the maximally mixed state (Remark~\ref{rem:gill_massar}), and the locate-then-measure protocol is the two stage adaptive scheme of quantum estimation theory~\cite{GillMassar2000,BarndorffNielsenGill2000,Fujiwara2006,HayashiMatsumoto2008}, where it yields a constant factor gain; here the hidden label amplifies that gain to an exponential one. Separately, the adaptive/nonadaptive comparison coincides with the pre- and post-measurement-information scenarios of quantum state discrimination, whose gap is known to witness measurement incompatibility~\cite{BallesterWehnerWinter2008,GopalWehner2010,CarmeliHeinosaariToigo2018,CarmeliHeinosaariToigo2019,HeinosaariMiyaderaZiman2016,SkrzypczykSupicCavalcanti2019,BuscemiChitambarZhou2020}; this is what lets us identify incompatibility as the resource behind the separation.

Taken together, these works leave open the central question addressed here: can a structured state family exhibit an exponential adaptive advantage even when the nonadaptive learner is allowed arbitrary global single copy POVMs? We show that the answer is yes, while also identifying why a simpler binary construction fails at this level of measurement power.

\subsection{Contributions and paper organization}

The contributions of this paper can be summarized as follows.

\begin{enumerate}[label=\textbf{C\arabic*.}]
    \item \textbf{A binary family and separation under product measurements} (Theorems~\ref{thm:binary_pauli_separation} and~\ref{thm:binary_product_povm_separation}).
    We introduce a binary hidden label family that isolates the adaptive localization and measurement mechanism described above. Under the local asymptotic minimax MSE criterion studied here, the adaptive protocol achieves an error of order \(1/N\), while nonadaptive product Pauli measurements exhibit the same \(1/N\) scaling with a leading constant that can be exponentially larger. We then show that the same exponential separation persists when the nonadaptive measurement class is enlarged to arbitrary product POVMs.

    \item \textbf{A precise explanation of why the binary family fails globally} (Theorem~\ref{thm:binary_global_no_separation}).
    We show that the support dependent signal observables in the binary family commute. Consequently, a fixed global single copy measurement can access all candidate signal statistics in a common eigenbasis and select the relevant one after the hidden label has been inferred. This gives an explicit mechanism by which a more powerful nonadaptive measurement eliminates the binary adaptive advantage.

    \item \textbf{A ternary family and separation under global measurements} (Theorems~\ref{thm:ternary_separation} and~\ref{thm:ternary_finite_sample}).
    Motivated by the preceding gap, we introduce a ternary hidden label construction and prove an aggregate Fisher information bound for arbitrary global POVMs acting on one copy at a time. This bound implies that every nonadaptive protocol has some hidden label for which the available Fisher information is exponentially small. By contrast, an adaptive protocol can first identify the hidden label from the marker register and then measure the corresponding signal observable. The local asymptotic minimax MSE analysis yields a \(2^{\Omega(n)}\) lower bound on the ratio between the nonadaptive and adaptive leading constants, where \(n\) is the number of qubits per copy. The same information budget gives a finite sample version of the separation: at fixed accuracy, a number of state copies logarithmic in \(n\) suffices adaptively, using single qubit Pauli measurements and one round of adaptation, whereas every nonadaptive protocol with arbitrary global single copy POVMs needs exponentially many copies.

    \item \textbf{A separation within a simple structured state class} (Corollary~\ref{cor:family_separability} and Proposition~\ref{prop:ternary_mpo}).
    We show that the relevant state families are fully separable (indeed classical, being diagonal in a label dependent product basis) and that, when each marker qubit is grouped with its corresponding signal qubit, the ternary family admits an exact MPO representation with constant bond dimension. The separation therefore occurs without entanglement in the unknown state and without collective measurements across copies.

    \item \textbf{The quantum nature of the separation} (Proposition~\ref{prop:commuting_simulation} and Theorem~\ref{thm:ternary_joint_budget}).
    The unknown states are classical: each is diagonal in a product basis that depends on the hidden label, so the separation owes nothing to entanglement or to any other nonclassical correlation inside a state. We show that no commuting family of states, classical or quantum, admits an adaptive advantage once each copy may be measured arbitrarily, so the global separation has no classical analogue, and we identify the responsible resource as measurement incompatibility: the candidate observables cannot be measured jointly, and a single measurement must divide a limited information budget among exponentially many of them, whereas an adaptive learner can commit to the right one.
\end{enumerate}

Throughout the paper, ``single copy'' means that different state copies are never measured collectively. A global single copy POVM may act jointly on all qubits within one copy, but coherent or collective measurements across several copies are beyond the scope of this paper.

The remainder of the paper follows the measurement hierarchy. Section~\ref{sec:framework} formalizes the state model, measurement classes, Fisher information, and local asymptotic minimax MSE criterion. Section~\ref{sec:binary_family} analyzes the binary family under product Pauli measurements and arbitrary product POVMs. Section~\ref{sec:binary_global_failure} shows why the binary separation disappears once arbitrary global measurements within each copy are permitted. Section~\ref{sec:ternary_global_separation} introduces the ternary family and establishes the aggregate Fisher information bottleneck, deriving the nonadaptive lower bound and adaptive upper bound that yield the exponential separation against arbitrary global single-copy POVMs, alongside a nonadaptive upper bound based on 2-design measurements. Section~\ref{sec:structural_properties} establishes the separability, classicality, and MPO structure of the state families, proves that the global separation has no classical analogue, and interprets the separation in terms of measurement incompatibility. Finally, Section~\ref{sec:discussion_conclusions} synthesizes the mechanism, scope, broader implications, and open directions.

\section{Statistical Model and Measurement Framework}
\label{sec:framework}

This section defines the statistical model, the measurement classes, and the performance criterion used throughout the paper. The state model is introduced first, followed by the measurement hierarchy, Fisher information, the distinction between adaptive and nonadaptive protocols, and the local asymptotic minimax MSE criterion. The binary and ternary constructions introduced later are two specific instances of this common framework.

\subsection{Marker--signal state model}
\label{sec:state_model}

We consider \(N\) independently prepared copies of an unknown state \(\rho_{\lambda,c}
=
\tau_{\lambda}\otimes\sigma_{\lambda,c}\). Each copy contains two registers: a marker register $\tau_{\lambda}$ and a signal register $\sigma_{\lambda,c}$, each consisting of \(L\) qubits. Thus each state copy contains \(n=2L\) qubits in total.

The discrete variable \( \lambda\in\Lambda_L\) is an unknown hidden label drawn from a finite label set \(\Lambda_L\). The scalar \( c\in\Theta\subseteq[-1,1] \) is the continuous parameter that we wish to estimate. Both the label \(\lambda\) and the parameter \(c\) are fixed across all \(N\) copies. In the language of quantum estimation theory, \(c\) is the parameter of interest and \(\lambda\) is a discrete nuisance parameter: it is not what we want to know, but it determines which measurement is informative about what we want to know. The task is thus single parameter quantum estimation with a nuisance parameter under single copy measurements, as discussed in Section~\ref{sec:related_work}.

The marker state \(\tau_\lambda\) on the marker register depends on the hidden label but not on \(c\). Its role is to carry information about \(\lambda\). The signal state on the signal register has the form
\begin{equation}
\sigma_{\lambda,c}
=
2^{-L}\bigl(\Id_{L}+cP_\lambda\bigr),
\label{eq:generic_signal_state}
\end{equation}
where \(\Id_{L}\) denotes the identity on the \(L\)-qubit signal register and \(P_\lambda \in \{I, X, Y, Z\}^{\otimes L} \setminus \{\Id_{L}\}\) is a non-identity \(L\)-qubit Pauli string determined by the hidden label. Since \(
P_\lambda\) is traceless and satisfies \( P_\lambda^2=\Id_{L} \), the signal state in \eqref{eq:generic_signal_state} has unit trace and is positive semidefinite for \(|c|\leq1\).

The two constructions studied later differ in their choices of the label set \(\Lambda_L\), the marker states \(\tau_\lambda\), and the Pauli strings \(P_\lambda\). Their common statistical structure is already visible at the reference point \(c=0\):
\begin{equation}
\sigma_{\lambda,0}
=
2^{-L}\Id_{L},
\qquad
\left.
\frac{\partial\sigma_{\lambda,c}}{\partial c}
\right|_{c=0}
=
2^{-L}P_\lambda.
\label{eq:base_and_tangent}
\end{equation}
Thus, the signal state itself is identical for every hidden label at \(c=0\), but a small change in \(c\) perturbs the state in a label dependent direction. The hidden label $\lambda$ therefore encodes the locally informative signal direction. The marker register provides information about which direction is relevant, while the signal register provides information about \(c\) once that direction is known.

\subsection{Single copy measurement hierarchy}
\label{sec:measurement_hierarchy}

A positive operator valued measure (POVM) on one state copy is a collection
\[
\mathcal M=\{E_y\}_{y\in\mathcal Y},
\qquad
E_y\succeq0,
\qquad
\sum_{y\in\mathcal Y}E_y=\Id_{2L},
\]
where the identity now acts on the full \(2L\)-qubit marker--signal Hilbert space. If \(\mathcal M\) is applied to \(\rho_{\lambda,c}\), the probability of outcome \(y\) is
\begin{equation}
p_y^{\mathcal M}(\lambda,c)
=
\Tr\!\left(E_y\rho_{\lambda,c}\right).
\label{eq:generic_outcome_probability}
\end{equation}

We consider three increasingly general single copy measurement classes.

\paragraph{Product Pauli measurements}
On the signal register, each qubit is measured separately in one of the Pauli bases \(X\), \(Y\), or \(Z\). A signal setting is therefore specified by \(
\mathbf b=(b_1,\ldots,b_L)\in\{X,Y,Z\}^L \), and, for outcomes \(
\mathbf r=(r_1,\ldots,r_L)\in\{+1,-1\}^L \), the corresponding signal POVM element is \(
E_{\mathbf r}^{(\mathbf b)}
=
\bigotimes_{\ell=1}^L
\frac{\Id_{1}+r_\ell b_\ell}{2} \), where $\Id_{1}$ here denotes the single-qubit identity. The marker qubits may simultaneously be read out in the \(Z\) basis.

\paragraph{Product POVMs}
The Pauli restriction on the signal register is removed. At signal site \(\ell\), one may use an arbitrary single qubit POVM \(
\mathcal M_\ell=\{E_{\ell,y_\ell}\}_{y_\ell} \), where \(y_\ell\) indexes the local measurement outcome at site \(\ell\). For a joint outcome $\mathbf{y} = (y_1, \dots, y_L)$, the signal POVM elements factor as
\begin{equation}
E_{\mathbf y}
=
\bigotimes_{\ell=1}^L E_{\ell,y_\ell}.
\label{eq:product_povm}
\end{equation}
Thus arbitrary single qubit POVMs are allowed at each site, but the joint signal measurement must remain a tensor product of single qubit measurements. As in the product Pauli model, the marker register is read out locally in the \(Z\) basis when marker information is used.

\paragraph{Global single copy POVMs}
The product restriction is removed entirely. A global single copy POVM may act jointly on all \(2L\) marker and signal qubits of each copy. The measurement may jointly couple different qubits, including qubits from both the marker and signal registers. In particular, this class includes informationally complete POVMs on the full \(2L\)-qubit system.

These measurement classes form the hierarchy
\[
\text{product Pauli}
\;\subset\;
\text{product POVM}
\;\subset\;
\text{global single copy POVM}.
\]
The distinction between the last class and a collective multi copy measurement is important. A global single copy POVM may jointly measure all qubits \emph{within one copy}, but every protocol considered in this paper acts on the \(N\) independently prepared copies one at a time. No POVM acts coherently or collectively on two or more different copies.

\subsection{Fisher information}
\label{sec:fisher_information}

For a fixed POVM \(\mathcal M\), we use the classical Fisher information of the resulting outcome distribution to quantify its local sensitivity to \(c\). Using \eqref{eq:generic_outcome_probability}, define
\begin{equation}
I_{\mathcal M}(\lambda;c)
=
\sum_{y:\,p_y^{\mathcal M}(\lambda,c)>0}
\frac{
\left[
\partial_c p_y^{\mathcal M}(\lambda,c)
\right]^2
}{
p_y^{\mathcal M}(\lambda,c)
}.
\label{eq:generic_fisher_information}
\end{equation}
A large value of \(I_{\mathcal M}(\lambda;c)\) means that the outcome distribution changes substantially under a small change in \(c\), whereas zero Fisher information means that the measurement is locally insensitive to \(c\) at that state.

For a predetermined sequence of measurements \(\mathcal M^{(1)},\ldots,\mathcal M^{(N)}\), the measurement outcomes from different independently prepared copies are independent, so their Fisher information adds:
\begin{equation}
J_\lambda(c)
=
\sum_{t=1}^N
I_{\mathcal M^{(t)}}(\lambda;c).
\label{eq:total_nonadaptive_fisher}
\end{equation}
For adaptive protocols, the measurement on each copy may depend on earlier outcomes, so its Fisher information is evaluated conditionally on the previous measurement history.

Unless stated otherwise, ``Fisher information'' in this paper refers to the classical Fisher information of a specified measurement, as in \eqref{eq:generic_fisher_information}. Quantum Fisher information is used only later as a measurement-independent upper bound on the information available from one copy.

\subsection{Adaptive and nonadaptive protocols}
\label{sec:adaptive_nonadaptive}

A protocol specifies which measurement is applied to each copy and how the resulting data are used to produce an estimate \(\widehat c_N\) of \(c\).

A \emph{nonadaptive} protocol chooses its entire sequence of measurements before observing any measurement outcomes. The sequence may be deterministic or may be randomized in advance, but any such randomization is independent of the unknown state and of all measurement outcomes. The realized measurement choices are known to the estimator and are part of the experimental record.

An \emph{adaptive} protocol may choose the measurement on copy \(t\) as a function of the outcomes observed on copies \( 1,\ldots,t-1 \). The measurement applied to copy \(t\) must therefore be selected before any outcome from copy \(t\) is observed. Adaptation occurs only between independently prepared copies.

The distinction concerns only how measurements are selected. After the data have been collected, both adaptive and nonadaptive protocols may use arbitrary classical post processing of the complete measurement record. In particular, a nonadaptive protocol is free to infer the hidden label \(\lambda\) from its data and then use that inferred label when constructing its final estimate; what it cannot do is use that information to change measurements that have already been fixed.

Measurement class and adaptivity are separate aspects of the protocol. Later sections state explicitly whether a comparison is made under product Pauli, product POVM, or global single copy measurements.

\subsection{Local asymptotic minimax MSE}
\label{sec:local_minimax}

The primary loss considered in this paper is the squared error in estimating the scalar parameter \(c\). For a protocol \(\mathsf P_N\) and estimator \(\widehat c_N\), define the pointwise MSE
\begin{equation}
\mathcal R_N
\bigl(
\lambda,c;
\mathsf P_N,\widehat c_N
\bigr)
:=
\E_{\lambda,c}^{\mathsf P_N}
\left[
(\widehat c_N-c)^2
\right].
\label{eq:pointwise_mse}
\end{equation}
The expectation is over the measurement outcomes and any randomness used by the protocol. The loss measures only the squared error in estimating \(c\); errors in identifying the hidden label \(\lambda\) matter only through their effect on the final estimate of \(c\).

Let \(\Pi_N\) denote a specified class of \(N\)-copy protocols, for example nonadaptive product POVMs or adaptive global single copy POVMs. For a reference value \(c_0\) and neighborhood width \(w>0\), define the finite-\(N\) local minimax MSE
\begin{equation}
R_{\Pi}
(N;c_0,w)
:=
\inf_{\mathsf P_N\in\Pi_N}
\inf_{\widehat c_N}
\;
\sup_{\lambda\in\Lambda_L}
\sup_{\substack{
c\in\Theta\\
|c-c_0|\leq w
}}
\mathcal R_N
\bigl(
\lambda,c;
\mathsf P_N,\widehat c_N
\bigr).
\label{eq:generic_local_minimax_mse}
\end{equation}
Thus the protocol and estimator are chosen first, after which performance is evaluated at the worst hidden label and the worst parameter value in the prescribed local neighborhood. No unbiasedness or regularity assumption is imposed on \(\widehat c_N\).

For the asymptotic results, fix \(
\epsilon\in\left(0,\frac12\right)
\) and choose
\begin{equation}
w_N
=
N^{-1/2+\epsilon}.
\label{eq:local_window}
\end{equation}
This window has the two properties
\begin{equation}
w_N\longrightarrow0,
\qquad
\sqrt N\,w_N
=
N^\epsilon
\longrightarrow\infty.
\label{eq:local_window_properties}
\end{equation}
The first property makes the criterion local around \(c_0\). The second makes the neighborhood wider than the canonical \(N^{-1/2}\) estimation scale.

This second condition is important because it prevents an estimator from appearing artificially accurate simply by being tuned to the single reference point \(c_0\). For example, the estimator \( \widehat c_N\equiv c_0 \) has worst case squared error \(w_N^2\) over the local window. Since \(Nw_N^2=N^{2\epsilon}\to\infty\), the worst case error of this estimator is asymptotically much larger than the \(1/N\) scale. By evaluating the worst case error over a shrinking neighborhood of \(c_0\), we therefore require an estimator to perform well locally around \(c_0\), rather than only at that single point.

The condition \( Nw_N^2\to\infty \) is also useful for the van Trees lower bounds used later. A smooth prior supported on this window can be chosen with Fisher information of order \(O(w_N^{-2})\), and because \(w_N=N^{-1/2+\epsilon}\), \( O(w_N^{-2})=o(N) \). Thus the Fisher information contributed by the prior is asymptotically smaller than the \(O(N)\) information contributed by the measurements.

We therefore abbreviate
\begin{equation}
R_{\Pi}(N;c_0)
:=
R_{\Pi}(N;c_0,w_N)
\label{eq:local_minimax_abbreviation}
\end{equation}
when the choice \eqref{eq:local_window} is understood. The normalized quantity \( N R_{\Pi}(N;c_0) \) is the object compared in our asymptotic results. When it converges to a finite value, that value is the leading constant multiplying the usual \(1/N\) local MSE; when an exact limit is unnecessary, the corresponding statements are made using \(\liminf\) or \(\limsup\).

The criterion in \eqref{eq:generic_local_minimax_mse} concerns the scalar estimation error in \(c\). Whenever a result is transferred to a distance between quantum states, that relation is stated separately rather than identifying scalar MSE with full state reconstruction error.

Finally, all structural parameters defining a particular state family are held fixed while \( N\to\infty \). Only after taking this large-\(N\) limit do we examine how the resulting asymptotic constants scale with the system size \(L\), or equivalently with the total qubit count \(n=2L\). Consequently, the exponential separations established later concern the system size dependence of these fixed-\(L\) asymptotic constants; they are not joint \(N,L\to\infty\) limits.

\section{Binary Family: Separation Under Product Measurements}
\label{sec:binary_family}

We now introduce a simple family that isolates the mechanism behind the adaptive advantage. A hidden support determines which signal observable contains information about the scalar parameter~\(c\). An adaptive protocol can first learn this support from the marker register and then measure the corresponding signal observable. A nonadaptive protocol must choose its signal measurements before the support is known and therefore has to distribute its measurement effort across many possible informative directions.

We first analyze this mechanism under product Pauli measurements and then ask whether the same separation persists when the local measurement class is enlarged to arbitrary product POVMs.

\subsection{Binary state family}
\label{sec:binary_state_family}

Fix integers \(1\leq K\leq L\), write \([L]:=\{1,\ldots,L\}\), and let
\begin{equation}
\mathcal S_{L,K}
=
\{S\subseteq[L]:|S|=K\},
\qquad
M_{L,K}
:=
|\mathcal S_{L,K}|
=
\binom{L}{K}.
\label{eq:binary_support_family}
\end{equation}
For the binary family, we identify the generic hidden label \(\lambda\) from Section~\ref{sec:state_model} with a support \(S\in\mathcal S_{L,K}\), and we use \(S\) throughout this section. We take the scalar parameter \(c\) to lie in
\[
\Theta=[-c_{\max},c_{\max}],
\qquad
0<c_{\max}<1,
\]
ensuring $c = 0$ is an interior point.

Each state copy contains \(L\) marker qubits and \(L\) signal qubits. For a known marker strength \(u\in(0,1]\), define
\begin{equation}
\tau_S
=
\bigotimes_{\ell=1}^{L}
\frac{1}{2}
\left(
\Id_{1}+\alpha_{\ell,S}Z
\right),
\qquad
\alpha_{\ell,S}
=
\begin{cases}
u, & \ell\in S,\\
0, & \ell\notin S.
\end{cases}
\label{eq:binary_marker_state}
\end{equation}
Thus measuring marker qubit \(\ell\) in the \(Z\) basis produces a \(\{\pm1\}\)-valued outcome with mean \(u\) when \(\ell\in S\) and mean \(0\) otherwise.

The hidden support also determines the signal Pauli string
\begin{equation}
P_S
=
\bigotimes_{\ell=1}^{L}A_{\ell,S},
\qquad
A_{\ell,S}
=
\begin{cases}
X, & \ell\in S,\\
Y, & \ell\notin S.
\end{cases}
\label{eq:binary_codeword}
\end{equation}
The signal state and full marker-signal state are
\begin{equation}
\sigma_{S,c}
=
2^{-L}
\left(
\Id_{L}+cP_S
\right),
\qquad
\rho_{S,c}
=
\tau_S\otimes\sigma_{S,c}.
\label{eq:binary_full_state}
\end{equation}
Because \(P_S^2=\Id_{L}\), the signal state is positive semidefinite for \(|c|\leq1\).

We refer to \eqref{eq:binary_marker_state}--\eqref{eq:binary_full_state} as the \emph{binary family}: at each site, membership or nonmembership in \(S\) selects between two marker values and, correspondingly, between the two signal directions \(X\) and \(Y\).

\subsection{Local geometry at \(c=0\)}
\label{sec:binary_geometry}

The binary family specializes the local structure described in Section~\ref{sec:state_model}. At \(c=0\), the signal state is \(2^{-L}\Id_{L}\) for every \(S\in\mathcal S_{L,K}\), while its derivative with respect to \(c\) is \(2^{-L}P_S\). Thus the \emph{signal} component of the state is identical for every support at \(c=0\), but a small change in \(c\) perturbs that signal state in a support-dependent direction. The full states \(\rho_{S,0}\) are not identical, because their marker states \(\tau_S\) still depend on \(S\). The marker register can therefore reveal the support and, with it, which of the \(M_{L,K}\) candidate signal observables is relevant for estimating \(c\).

\subsection{Adaptive support localization}
\label{sec:binary_localization}

The adaptive protocol first localizes the hidden support using the marker register. During this localization stage, the signal register is measured using any fixed product Pauli setting and those signal outcomes are discarded. The following lemma specifies the localization procedure and its error guarantee; its proof is given in Appendix~\ref{app:binary_pauli}.

\begin{lemma}[Top-\(K\) marker localization]
\label{lem:binary_marker_localization}
Fix \(L\geq1\), \(1\leq K\leq L\), and marker strength \(u>0\). Use \(m_0\) independently prepared copies of the full state. On each of these \(m_0\) copies, measure all \(L\) marker qubits in the \(Z\) basis. Let
\[
Z_{\ell,t}\in\{+1,-1\},
\qquad
\ell\in[L],
\qquad
t\in[m_0],
\]
denote the outcome at marker site \(\ell\) on localization copy \(t\), and define \( \widehat\alpha_\ell
=
\frac1{m_0}
\sum_{t=1}^{m_0} Z_{\ell,t} \). Let \(\widehat S\) be the set of the \(K\) largest values among
\(\widehat\alpha_1,\ldots,\widehat\alpha_L\), with any fixed deterministic tie-breaking rule. Then, uniformly over all \(S\in\mathcal S_{L,K}\),
\begin{equation}
\Pr_S(\widehat S\neq S)
\leq
2L\exp\!\left(-\frac{m_0u^2}{8}\right).
\label{eq:binary_marker_error}
\end{equation}
In particular, choosing
\begin{equation}
m_0(N)
=
\left\lceil
\frac{8}{u^2}
\left[
\log(2L)+2\log(N+2)
\right]
\right\rceil
\label{eq:binary_marker_budget}
\end{equation}
gives
\begin{equation}
\sup_{S\in\mathcal S_{L,K}}
\Pr_S(\widehat S\neq S)
\leq
(N+2)^{-2}
=
o(N^{-1}),
\label{eq:binary_marker_error_rate}
\end{equation}
while, for fixed \(L,K,u\),
\begin{equation}
m_0(N)=O(\log N)=o(N).
\label{eq:binary_marker_cost}
\end{equation}
\end{lemma}

After localization, the remaining \(N-m_0\) copies are used to measure the signal observable \(P_{\widehat S}\). On these copies, the marker qubits are measured in the \(Z\) basis and their outcomes are discarded, since they are not needed for estimating \(c\). Operationally, this means measuring \(X\) on the signal qubits indexed by \(\widehat S\), measuring \(Y\) on the remaining signal qubits, and multiplying the resulting \(\{\pm1\}\) outcomes to form a parity statistic.

If \(\widehat S=S\), each such parity outcome \(R_t\in\{\pm1\}\) satisfies
\begin{equation}
\E_{S,c}[R_t]=c,
\qquad
\operatorname{Var}_{S,c}(R_t)=1-c^2.
\label{eq:binary_matched_parity}
\end{equation}
The adaptive estimate of \(c\) is the sample mean of these parity outcomes, clipped to the physical interval~\(\Theta\). Theorem~\ref{thm:binary_pauli_separation} below fully characterizes the performance of the adaptive protocol.

\subsection{Product Pauli benchmark}
\label{sec:binary_product_pauli}

We first restrict the signal measurements to product Pauli settings. A setting \( \mathbf b=(b_1,\ldots,b_L)\in\{X,Y,Z\}^L \) matches support \(S\) when \( b_\ell=A_{\ell,S}\) for every \(\ell\).

The following observation explains the combinatorial cost of a nonadaptive design. Its proof, together with the proof of the separation theorem in this subsection, is given in Appendix~\ref{app:binary_pauli}.

\begin{lemma}[Only the matched Pauli setting is informative]
\label{lem:binary_pauli_information}
For a product-Pauli measurement on the signal register,
\[
I_{\mathbf b}(S;c)
=
\begin{cases}
\dfrac{1}{1-c^2},
&
\mathbf b
=
(A_{1,S},\ldots,A_{L,S}),
\\[2mm]
0,
&
\text{otherwise}.
\end{cases}
\]
In particular, any measurement setting with an identity factor, or with even one mismatched local Pauli axis, gives no information about \(c\).
\end{lemma}

To convert a uniform Fisher information bound into a local asymptotic minimax MSE lower bound, we use Lemma~\ref{lem:local_fisher_to_mse}, stated and proved in Appendix~\ref{app:fisher_mse}. It applies without an unbiasedness assumption and will be used in the lower bounds below.

Let \( R_{\mathrm{AD,PP}}^{(L,K)}(N;c_0) \) and \( R_{\mathrm{NA,PP}}^{(L,K)}(N;c_0) \) denote the local asymptotic minimax MSE
\eqref{eq:generic_local_minimax_mse}, with \(w=w_N\), for adaptive and nonadaptive product Pauli protocols, respectively. We quantify and compare these MSEs in the following theorem.

\begin{theorem}[Separation under product Pauli measurements]
\label{thm:binary_pauli_separation}
For every fixed \(L,K,u\) and every \(c_0\in(-c_{\max},c_{\max})\), 
\begin{equation}
\lim_{N\to\infty}
N
R_{\mathrm{AD,PP}}^{(L,K)}(N;c_0)
=
1-c_0^2,
\label{eq:binary_adaptive_constant}
\end{equation}
whereas
\begin{equation}
\liminf_{N\to\infty}
N
R_{\mathrm{NA,PP}}^{(L,K)}(N;c_0)
\geq
M_{L,K}(1-c_0^2)
=
\binom{L}{K}(1-c_0^2).
\label{eq:binary_nonadaptive_pauli_constant}
\end{equation}
Thus the adaptive and nonadaptive \(1/N\)-scale MSEs differ by at least a factor \( \binom{L}{K} \).
\end{theorem}

Key to Theorem~\ref{thm:binary_pauli_separation} is the combinatorial number of possible supports. Once \(S\) has been identified, the adaptive protocol can spend all remaining copies measuring \(P_S\). A nonadaptive protocol cannot know in advance which of the \( M_{L,K}=\binom{L}{K} \) codewords will be relevant, and some support must therefore receive at most a \(1/M_{L,K}\) share of its matched measurements.

If \(K\) is fixed, this produces a polynomial separation, as \( \binom{L}{K}
=
\Theta(L^K) \). However, if
\(
K=\lfloor\alpha L\rfloor\) for
\(0<\alpha<1\),
then the separation is exponential, as
\begin{equation}
\binom{L}{\lfloor\alpha L\rfloor}
=
\frac{
2^{L H_2(\alpha)}
}{
\Theta(\sqrt L)
}
=
2^{\Theta(L)}
=
2^{\Theta(n)},
\label{eq:binary_exponential_regime}
\end{equation}
where
\(
H_2(\alpha)
=
-\alpha\log_2\alpha
-
(1-\alpha)\log_2(1-\alpha)
\)
is the binary entropy function and \(n=2L\) is the number of qubits per copy.

\subsection{Extension to arbitrary product POVMs}
\label{sec:binary_product_povm}

The preceding result leaves open the possibility that the separation occurs only because the nonadaptive protocol is forced to choose among the Pauli bases. We now remove that restriction and allow an arbitrary single qubit POVM at every signal site, while retaining the product structure across sites.

The product POVM lower bound rests on a sequence of one-copy information bounds collected in Appendix~\ref{app:binary_product_povm}. Lemma~\ref{lem:binary_local_xy_budget} bounds the combined local sensitivity to the \(X\) and \(Y\) directions, while Lemma~\ref{lem:binary_product_factorization} factorizes the Fisher information across signal sites at \(c=0\), where the common signal state is maximally mixed. Lemma~\ref{lem:binary_support_fisher_budget} then shows that the Fisher information summed over all hidden supports is at most one. Lemma~\ref{lem:binary_product_nearzero} extends our analysis from \(c=0\) to the shrinking local neighborhood.

The central consequence is that a fixed product POVM may distribute its information unevenly across the \(\binom{L}{K}\) possible supports, but it cannot provide order one Fisher information for all of them simultaneously. These bounds lead to the same separation guarantee as in the product Pauli setting.

Let \( R_{\mathrm{AD,Prod}}^{(L,K)}(N;0) \) and \( R_{\mathrm{NA,Prod}}^{(L,K)}(N;0) \) denote the local asymptotic minimax MSEs at \(c_0=0\) for adaptive and nonadaptive product POVM protocols, respectively. We quantify and compare these MSEs in the following theorem. Its proof is also given in Appendix~\ref{app:binary_product_povm}.

\begin{theorem}[Separation under arbitrary product POVMs]
\label{thm:binary_product_povm_separation}
For fixed \(L,K,u\),
\begin{equation}
\lim_{N\to\infty}
N
R_{\mathrm{AD,Prod}}^{(L,K)}(N;0)
=
1,
\label{eq:binary_product_adaptive}
\end{equation}
whereas
\begin{equation}
\liminf_{N\to\infty}
N
R_{\mathrm{NA,Prod}}^{(L,K)}(N;0)
\geq
M_{L,K}
=
\binom{L}{K}.
\label{eq:binary_product_nonadaptive}
\end{equation}
Consequently, the adaptive and nonadaptive \(1/N\)-scale local MSEs differ by at least a factor \( \binom{L}{K} \). For
\(K=\lfloor\alpha L\rfloor\) with
\(0<\alpha<1\), this factor is \( 2^{\Theta(L)}
=
2^{\Theta(n)} \).
\end{theorem}

Theorem~\ref{thm:binary_product_povm_separation} shows that for the binary family, the exponential MSE separation is not simply an artifact of restricting the nonadaptive learner to Pauli bases. The remaining question is whether the \emph{product} measurement restriction is responsible for this bottleneck. We therefore next consider global single copy measurements that can act jointly on all qubits within each state copy.

\section{Nonadaptive Global Measurements on the Binary Family}
\label{sec:binary_global_failure}

We now allow arbitrary global single copy POVMs, as defined in Section~\ref{sec:measurement_hierarchy}. Such a measurement may act jointly on all marker and signal qubits within each state copy, but measurements on different copies remain separate. For the binary family, we see that we lose the exponential separation between adaptive and nonadaptive protocols. The reason is structural: although the informative signal direction depends on the hidden support, all candidate signal observables can be measured simultaneously in one common basis.

\subsection{Commutativity and a common signal measurement}

Recall that \(P_S=\bigotimes_{\ell=1}^L A_{\ell,S}\), where \(A_{\ell,S}=X\) for \(\ell\in S\) and \(A_{\ell,S}=Y\) otherwise. Consider two supports \(S,T\in\mathcal S_{L,K}\). The local Pauli factors of \(P_S\) and \(P_T\) differ exactly on the symmetric difference \(S\triangle T\). Since both supports have cardinality \(K\), we have \( |S\triangle T| = 2\bigl(K-|S\cap T|\bigr) \) which is always even. Each mismatched site contributes one anticommutation sign between \(X\) and \(Y\), and therefore \( P_S P_T = (-1)^{|S\triangle T|} P_T P_S = P_T P_S \). Thus the entire family \(\{P_S:S\in\mathcal S_{L,K}\}\) consists of mutually commuting Hermitian operators and admits a common orthonormal eigenbasis on the signal register.

Fix a common orthonormal eigenbasis \( \{|z\rangle\}_{z=1}^{2^L} \) of the commuting family \( \{P_S:S\in\mathcal S_{L,K}\} \) on the signal register. For every support \(S\), let \(\xi_S(z)\in\{+1,-1\}\) denote the corresponding eigenvalue, so that \( P_S|z\rangle = \xi_S(z)|z\rangle \). A projective measurement in this basis is fixed in advance and does not depend on the unknown support. Nevertheless, a single outcome \(z\) determines \(\xi_S(z)\) for every candidate support \(S\) simultaneously.
This is precisely the capability that was unavailable to product measurements in the preceding section.

Indeed, if the true support is \(S\), then for any \(T\in\mathcal S_{L,K}\),
\begin{equation}
\E_{S,c}\!\left[\xi_T(z)\right]
=
\Tr\!\left(P_T\sigma_{S,c}\right)
=
\begin{cases}
c, & T=S,\\
0, & T\neq S.
\end{cases}
\label{eq:binary_global_candidate_statistics}
\end{equation}
Thus the outcome of one fixed global signal measurement contains a stored statistic for every candidate support. The hidden support need not be known when the measurement is performed.

\subsection{A fixed nonadaptive global protocol}

The preceding observation yields an explicit nonadaptive protocol. On every copy, measure all marker qubits in the \(Z\) basis and measure the signal register in the fixed common eigenbasis above. This same measurement is used on every copy and is chosen before any data are observed. 
After all \(N\) copies have been measured, use the marker outcomes to form the support estimate \(\widehat S\) from Lemma~\ref{lem:binary_marker_localization}. Here the marker estimator may use all \(N\) copies, so the lemma with \(m_0=N\) gives
\[
\sup_{S\in\mathcal S_{L,K}}
\Pr_S(\widehat S\neq S)
\leq
2L\exp\!\left(-\frac{Nu^2}{8}\right)
=
o(N^{-1}).
\]
No signal measurement needs to be changed after \(\widehat S\) is obtained. For the previously recorded signal outcome \(z_t\) on copy \(t\), simply select the stored eigenvalue \( V_t := \xi_{\widehat S}(z_t) \). The estimate of \(c\) is the sample mean of \(V_1,\ldots,V_N\), clipped to the physical interval \(\Theta\).

Let \(R_{\mathrm{AD,Glob}}^{(L,K)}(N;c_0)\) and \(R_{\mathrm{NA,Glob}}^{(L,K)}(N;c_0)\) denote the local asymptotic minimax MSEs for adaptive and nonadaptive arbitrary global single copy POVM protocols, respectively. The proof of the theorem below is given in Appendix~\ref{app:binary_global}.

\begin{theorem}[No separation under global POVMs]
\label{thm:binary_global_no_separation}
For every fixed \(L,K,u\) and every \(c_0\in(-c_{\max},c_{\max})\),
\begin{equation}
\lim_{N\to\infty}
N
R_{\mathrm{AD,Glob}}^{(L,K)}(N;c_0)
=
\lim_{N\to\infty}
N
R_{\mathrm{NA,Glob}}^{(L,K)}(N;c_0)
=
1-c_0^2.
\label{eq:binary_global_equal_constants}
\end{equation}
Hence the binary family has no asymptotic adaptive advantage once arbitrary global single copy POVMs are permitted.
\end{theorem}

The failure to maintain the MSE separation for the binary family under global measurements reveals a limitation of the mechanism developed in the preceding section. Hiding the informative direction is not sufficient by itself. If all candidate directions are jointly measurable, a nonadaptive protocol can measure them simultaneously and postpone the identification of the relevant direction to classical post processing.

The distinction is therefore not merely that the candidate tangent directions are different. What matters at the global level is whether one fixed measurement can access all of them simultaneously. For the binary family, commutativity provides such an opportunity. 

To maintain an adaptive separation against arbitrary global single copy POVMs, we therefore need a family whose label dependent signal observables do not admit this simultaneous measurement strategy. The next section introduces a ternary construction designed to remove this obstruction.

\section{Ternary Family: Separation Under Global Measurements}
\label{sec:ternary_global_separation}

Section~\ref{sec:binary_global_failure} identifies a limitation in the binary family: its support-dependent signal observables commute, so one fixed global measurement can retain the statistics associated with every candidate support. We now modify the family construction so that this common measurement opportunity is no longer available. The resulting ternary family yields an exponential gap between adaptive and nonadaptive protocols even when the latter may use arbitrary global single copy POVMs.

\subsection{Ternary state family and main result}
\label{sec:ternary_family}

Fix \(L\geq1\) and a marker strength \(0<u<\tfrac12\). For the ternary family, we identify the generic hidden label~\(\lambda\) from Section~\ref{sec:state_model} with a word \( \mathbf a=(a_1,\ldots,a_L)\in\{X,Y,Z\}^L \), and we use \(\mathbf a\) throughout this section. For \(a\in\{X,Y,Z\}\), define the single qubit marker states
\[
\tau_X=\frac12(\Id_{1}-uZ),
\qquad
\tau_Y=\frac12\Id_{1},
\qquad
\tau_Z=\frac12(\Id_{1}+uZ).
\]
For a label \(\mathbf a \in\{X,Y,Z\}^L\), set
\[
\tau_{\mathbf a}
=
\bigotimes_{\ell=1}^L\tau_{a_\ell},
\qquad
P_{\mathbf a}
=
\bigotimes_{\ell=1}^L a_\ell,
\]
and, for this family, take \(\Theta=[-1,1]\) and define
\begin{equation}
\rho_{\mathbf a,c}
=
\tau_{\mathbf a}\otimes\sigma_{\mathbf a,c},
\qquad
\sigma_{\mathbf a,c}
=
2^{-L}(\Id_{L}+cP_{\mathbf a}),
\qquad
c\in[-1,1].
\label{eq:ternary_state_family}
\end{equation}
There are \(3^L\) possible hidden labels, and each state contains \(n=2L\) qubits in total.

While the binary construction allowed a common signal eigenbasis for all candidate observables, the ternary family contains noncommuting Pauli strings; for example, two labels that differ only by replacing one \(X\) with \(Y\) produce anticommuting signal observables. Thus the entire family \(\{P_{\mathbf a}\}\) cannot share a common eigenbasis. The lower bound below uses a stronger quantitative fact: the \(3^L\) operators \(P_{\mathbf a}\) are distinct, Hilbert--Schmidt orthogonal Pauli strings, so one fixed measurement has a limited aggregate Fisher information budget across them.

Let \(R_{\mathrm{AD,Glob}}^{(L)}(N;0)\) and \(R_{\mathrm{NA,Glob}}^{(L)}(N;0)\) denote the local asymptotic minimax MSEs from \eqref{eq:generic_local_minimax_mse}--\eqref{eq:local_minimax_abbreviation}, with hidden label set \(\{X,Y,Z\}^L\), for adaptive and nonadaptive global single copy POVM protocols, respectively. Define
\begin{equation}
B_L
:=
\frac{(2^L-1)(1+u)^L}{3^L}
=
(1-2^{-L})
\left(\frac{2(1+u)}{3}\right)^L.
\label{eq:ternary_BL}
\end{equation}
Because \(u<\tfrac12\), the factor \(2(1+u)/3\) is strictly smaller than one. We quantify and compare the adaptive and nonadaptive MSEs in the following theorem.

\begin{theorem}[Exponential separation against arbitrary global single copy POVMs]
\label{thm:ternary_separation}
For every fixed \(L\geq1\) and \(0<u<\tfrac12\),
\begin{equation}
\lim_{N\to\infty}
N R_{\mathrm{AD,Glob}}^{(L)}(N;0)
=
1,
\label{eq:ternary_adaptive_constant}
\end{equation}
whereas
\begin{equation}
\liminf_{N\to\infty}
N R_{\mathrm{NA,Glob}}^{(L)}(N;0)
\geq
\frac{1}{B_L}
=
\frac{3^L}{(2^L-1)(1+u)^L}.
\label{eq:ternary_nonadaptive_constant}
\end{equation}
Consequently, the ratio between the nonadaptive lower asymptotic constant and the adaptive asymptotic constant is at least
\[
\frac{1}{B_L}
=
\Omega\!\left(
\left[\frac{3}{2(1+u)}\right]^L
\right)
=
2^{\Omega(n)},
\qquad
n=2L.
\]
\end{theorem}
Proposition~\ref{prop:ternary_nonadaptive_upper} below exhibits a nonadaptive protocol with \(\limsup_{N\to\infty}NR^{(L)}_{\mathrm{NA,Glob}}(N;0)\leq2^L+1\), so the nonadaptive/adaptive ratio lies between \(1/B_L\) and \(2^L+1\). Both bounds are exponential in \(n\), with bases \(3/(2(1+u))\) and \(2\) per signal qubit, respectively. Theorem~\ref{thm:ternary_finite_sample} establishes the corresponding finite-sample separation at fixed accuracy.

We prove Theorem~\ref{thm:ternary_separation} in the following subsections, beginning with the adaptive statement and then establishing the nonadaptive lower bound.

\subsection{Adaptive benchmark}
\label{sec:ternary_adaptive}

The adaptive protocol follows the same locate-then-measure principle as in the binary family. During localization, the marker qubits are measured in the \(Z\) basis; the signal register is measured using any fixed product Pauli setting and those signal outcomes are discarded. The marker outcomes distinguish the three possible local labels through their means \(-u\), \(0\), and \(+u\). Proofs of the localization lemma and the two adaptive bounds in this subsection are collected in Appendix~\ref{app:ternary_adaptive}.

\begin{lemma}[Ternary marker localization]
\label{lem:ternary_localization}
Measure every marker qubit in the \(Z\) basis on \(m_0\) independently prepared copies. For each site \(\ell\), let \(\bar Z_\ell\) be the sample mean of its \(\{\pm1\}\)-valued outcomes and set \(\widehat a_\ell=X,Y,Z\) according to whether \(\bar Z_\ell\) is below \(-u/2\), lies between \(-u/2\) and \(u/2\), or is above \(u/2\), respectively. Then \( \Pr_{\mathbf a}
\bigl(
\widehat{\mathbf a}\neq\mathbf a
\bigr)
\leq
2L\exp\!\left(-\frac{m_0u^2}{8}\right) \). Choosing \( m_0(N)
=
\left\lceil
\frac{8}{u^2}
\left[
\log(2L)+2\log(N+2)
\right]
\right\rceil \) gives
\[
\Pr_{\mathbf a}
\bigl(
\widehat{\mathbf a}\neq\mathbf a
\bigr)
\leq
(N+2)^{-2}
=
o(N^{-1}),
\qquad
m_0=O(\log N)=o(N),
\]
for fixed \(L\) and \(u\).
\end{lemma}

After localization, the remaining \(N-m_0\) copies are used to estimate \(c\). On each such copy, the marker qubits are measured in the \(Z\) basis and their outcomes are discarded. The signal qubits are measured in the local Pauli bases specified by \(P_{\widehat{\mathbf a}}\), and the product of the resulting \(\{\pm1\}\) outcomes gives one parity sample. The final estimate is the parity sample mean, clipped to \([-1,1]\).

\begin{theorem}[Adaptive achievability]
\label{thm:ternary_adaptive}
For the ternary family~\eqref{eq:ternary_state_family}, the two-stage adaptive product Pauli protocol (marker localization followed by measurement of \(P_{\widehat{\mathbf a}}\)) yields
\[
\limsup_{N\to\infty}
N R_{\mathrm{AD,Glob}}^{(L)}(N;0)
\leq
1.
\]
\end{theorem}

\begin{theorem}[Matching adaptive lower bound]
\label{thm:ternary_adaptive_lower}
For the ternary family~\eqref{eq:ternary_state_family}, the local asymptotic minimax MSE over all adaptive global single copy POVM protocols satisfies
\[
\liminf_{N\to\infty}
N R_{\mathrm{AD,Glob}}^{(L)}(N;0)
\geq
1.
\]
\end{theorem}

Combining Theorems~\ref{thm:ternary_adaptive} and \ref{thm:ternary_adaptive_lower} proves \eqref{eq:ternary_adaptive_constant}.

\subsection{Nonadaptive lower bound}
\label{sec:ternary_nonadaptive_roadmap}

The nonadaptive argument has five steps. First, we bound the total Fisher information that any fixed signal measurement can distribute across the \(3^L\) candidate Pauli observables. Second, we show that access to the marker register can increase this aggregate information by only a controlled factor. Third, averaging then identifies a hidden label with small Fisher information. Fourth, we extend the bound at \(c=0\) to the shrinking local window \(|c|\leq w_N\). Finally, we convert this information bottleneck into a local asymptotic minimax MSE lower bound.

\subsubsection{Signal-only Fisher information budget}
\label{sec:ternary_signal_budget}

Let $d = 2^L$ denote the dimension of the signal Hilbert space. We first isolate the information constraint on the signal register alone.

The proof of the following budget uses the Pauli--Parseval identity. Both that identity, stated as Lemma~\ref{lem:pauli_parseval}, and the proof of the budget are given in Appendix~\ref{app:ternary_fisher_budget}.

When a POVM acts on the signal register alone, we write \(I_{\mathcal F}(\mathbf a;c)\) for the Fisher information about \(c\) generated by the signal state \(\sigma_{\mathbf a,c}\).

\begin{lemma}[Signal-only global Fisher budget]
\label{lem:ternary_signal_budget}
For every POVM \(\mathcal F=\{F_y\}_y\) on the signal register,
\[
\sum_{\mathbf a\in\{X,Y,Z\}^L}
I_{\mathcal F}(\mathbf a;0)
\leq
d-1.
\]
\end{lemma}

Thus a fixed signal measurement has a finite aggregate information budget: it cannot provide order-one Fisher information simultaneously for all \(3^L\) orthogonal candidate Pauli observables.

\begin{remark}[Relation to the Gill--Massar inequality]
\label{rem:gill_massar}
Lemma~\ref{lem:ternary_signal_budget} is a special case of the Gill--Massar inequality~\cite{GillMassar2000}: for any POVM \(\mathcal F\) on a \(d\)-dimensional system and any regular parametrization \(\theta\mapsto\rho_\theta\) of the full state space at a full rank state,
\[
\Tr\bigl(H(\theta)^{-1}F_{\mathcal F}(\theta)\bigr)\leq d-1,
\]
where \(H\) is the symmetric logarithmic derivative quantum Fisher information matrix and \(F_{\mathcal F}\) is the classical Fisher information matrix of the POVM. In the Pauli coordinates \(\rho_\theta=d^{-1}\bigl(\Id_{L}+\sum_{j\geq1}\theta_jP_j\bigr)\) at \(\theta=0\), the symmetric logarithmic derivatives are the Pauli strings themselves and \(H(0)=\Id_{d^2-1}\), so the left-hand side reduces to \(\sum_{j\geq1}I_{\mathcal F}(j;0)\). Restricting this sum to the ternary Pauli strings gives the lemma. The proof in Appendix~\ref{app:ternary_fisher_budget} also shows when the full Pauli budget is saturated: for \(F_y\neq0\), equality in \(\Tr(F_y^2)\leq\Tr(F_y)^2\) holds exactly when \(F_y\) has rank one, and Proposition~\ref{prop:ternary_nonadaptive_upper} shows that a rank one, 2-design POVM distributes that budget uniformly, giving every non-identity Pauli string exactly \(1/(d+1)\). The underlying bound is thus not specific to the ternary family; it bounds how any fixed measurement can share information among Hilbert--Schmidt orthogonal tangent directions at the maximally mixed state.
\end{remark}

\subsubsection{Marker side information and the joint global Fisher budget}
\label{sec:ternary_joint_budget}

We next allow an arbitrary global measurement acting jointly on the marker and signal registers. The marker can reveal information about the hidden label, but the following result shows that this side information increases the aggregate Fisher information by only a controlled factor.

The marker contribution is quantified by Lemma~\ref{lem:ternary_marker_factor} in Appendix~\ref{app:ternary_fisher_budget}, which shows that \( \sum_z\max_{\mathbf a}p(z\mid\mathbf a)=(1+u)^L \). Together with the signal-only budget, this yields the joint bound below. Its proof is also given in Appendix~\ref{app:ternary_fisher_budget}.

\begin{theorem}[Joint marker--signal global Fisher budget]
\label{thm:ternary_joint_budget}
For every global single copy POVM \(\mathcal M=\{E_y\}_y\),
\begin{equation}
\sum_{\mathbf a\in\{X,Y,Z\}^L}
I_{\mathcal M}(\mathbf a;0)
\leq
(2^L-1)(1+u)^L.
\label{eq:ternary_joint_budget}
\end{equation}
\end{theorem}

Averaging Theorem~\ref{thm:ternary_joint_budget} over the \(3^L\) hidden labels gives Corollary~\ref{cor:ternary_worst_word} in Appendix~\ref{app:ternary_fisher_budget}: every global single copy POVM has at least one hidden label \(\mathbf a^\star\) for which
\(
I_{\mathcal M}(\mathbf a^\star;0)\leq B_L
\).

\begin{remark}[Classical and quantum parts of the budget]
\label{rem:budget_decomposition}
The two factors in \(3^LB_L=(2^L-1)(1+u)^L\) have different origins. The factor \((1+u)^L\) is classical: it is the maximum aggregate budget available to a learner who, within each copy, first reads the marker string \(z\) and then commits to one candidate signal observable chosen as a function of \(z\), since the Fisher information of this strategy satisfies
\[
\sum_{\mathbf a}I(\mathbf a;0)=\sum_z\sum_{\mathbf a}p(z\mid\mathbf a)\Pr[\text{choice}=\mathbf a\mid z]\leq\sum_z\max_{\mathbf a}p(z\mid\mathbf a)=(1+u)^L.
\]
The factor \(2^L-1\) is quantum: it is the Gill--Massar budget of Remark~\ref{rem:gill_massar}, which a single POVM can spread over all \(4^L-1\) Pauli directions at once, whereas a nonadaptive strategy that fixes one of the \(3^L\) candidate observables before observing each copy has aggregate budget \(1\). Thus, for \(L\geq2\), the 2-design protocol of Proposition~\ref{prop:ternary_nonadaptive_upper} has smaller worst-case asymptotic MSE at \(c=0\) than the restricted classical model of Proposition~\ref{prop:classical_restricted}, yet Theorem~\ref{thm:ternary_separation} shows that nonadaptive global protocols remain exponentially weaker than adaptive ones.
\end{remark}

\subsubsection{From \(c=0\) to the local window}
\label{sec:ternary_nearzero}

The preceding budget is pointwise at \(c=0\), whereas the local asymptotic minimax MSE is evaluated over the shrinking window \(|c|\leq w_N\). Lemma~\ref{lem:ternary_nearzero}, stated and proved in Appendix~\ref{app:ternary_nonadaptive}, provides the required uniform extension:
\[
I_{\mathcal M}(\mathbf a;c)
\leq
\frac{I_{\mathcal M}(\mathbf a;0)}{1-|c|}.
\]
Since \(w_N\to0\), this changes the information bound only by a \(1+o(1)\) factor over the local window.

\subsubsection{Worst hidden label and the nonadaptive lower bound}
\label{sec:ternary_nonadaptive}

Summing the one-copy joint budget over a nonadaptive \(N\)-copy measurement schedule and averaging over the hidden labels gives Lemma~\ref{lem:ternary_worst_word_N} in Appendix~\ref{app:ternary_nonadaptive}. It identifies a label \(\mathbf a_N^\star\), chosen before any data are observed, whose total Fisher information at \(c=0\) is at most \(N B_L\). The same conclusion holds for measurement schedules randomized in advance, after averaging over the recorded random schedule. The resulting nonadaptive lower bound is provided by Theorem~\ref{thm:ternary_nonadaptive}, which is proved in Appendix~\ref{app:ternary_nonadaptive}.

\begin{theorem}[Nonadaptive lower bound against global single copy POVMs]
\label{thm:ternary_nonadaptive}
For fixed \(L\geq1\) and \(0<u<\tfrac12\),
\[
\liminf_{N\to\infty}
N R_{\mathrm{NA,Glob}}^{(L)}(N;0)
\geq
\frac{1}{B_L}.
\]
\end{theorem}

Theorem~\ref{thm:ternary_nonadaptive} proves \eqref{eq:ternary_nonadaptive_constant}. Together with the adaptive benchmark, this completes the proof of Theorem~\ref{thm:ternary_separation}.

The ternary construction therefore removes the common eigenbasis measurement opportunity identified in Section~\ref{sec:binary_global_failure}. The exponential separation now persists even with arbitrary global single copy measurements within each independently prepared copy. Section~\ref{sec:structural_properties} examines the structural properties of the two constructions and shows that this phenomenon does not require entanglement in the unknown state, nor any nonclassical correlation in it, and that it has no classical analogue.

\subsection{Finite sample separation at fixed accuracy}
\label{sec:ternary_finite_sample}

For a measurement class \(\mathcal C\), let \(N_{\mathcal C}(L,u,\epsilon)\) be the least number of copies for which some protocol in \(\mathcal C\) and some estimator \(\widehat c\) achieve
\begin{equation}
\sup_{\mathbf a\in\{X,Y,Z\}^L}\ \sup_{c\in\Theta}
\Pr_{\mathbf a,c}\bigl(|\widehat c-c|>\epsilon\bigr)\leq\tfrac13 .
\label{eq:ternary_finite_accuracy}
\end{equation}
The protocol may depend on \(L\), \(u\), and \(\epsilon\) but not on \(\mathbf a\) or \(c\). We compare the adaptive product Pauli class \(\mathrm{AD,PP}\) with the nonadaptive global class \(\mathrm{NA,Glob}\), the latter allowing schedules randomized in advance. The proof of the theorem is given in Appendix~\ref{app:ternary_finite_sample}.

\begin{theorem}[Finite sample separation]
\label{thm:ternary_finite_sample}
For every \(L\geq1\), \(0<u<\tfrac12\), and \(0<\epsilon\leq\tfrac14\),
\begin{align}
N_{\mathrm{AD,PP}}(L,u,\epsilon)
&\leq
\left\lceil\frac{8}{u^2}\log(12L)\right\rceil+\left\lceil\frac{2}{\epsilon^2}\log12\right\rceil,
\label{eq:ternary_finite_adaptive}\\
N_{\mathrm{NA,Glob}}(L,u,\epsilon)
&\geq
\frac{1}{144\,B_L\,\epsilon^2},
\label{eq:ternary_finite_nonadaptive}
\end{align}
with \(B_L\) as in \eqref{eq:ternary_BL}. In particular, at fixed \(u\) and \(\epsilon\), adaptive product Pauli measurements need \(O(\log L)=O(\log n)\) copies, whereas nonadaptive global single copy measurements need \(2^{\Omega(n)}\) copies.
\end{theorem}

The lower bound applies to every estimator, biased or not, and allows arbitrary classical post processing. It is already obtained by comparing the two values \(c=\pm2\epsilon\) for a single hidden label. Theorem~\ref{thm:ternary_finite_sample} is a finite sample counterpart to Theorem~\ref{thm:ternary_separation}: the adaptive protocol uses Pauli measurements on individual qubits, one round of adaptation, and a number of copies that is logarithmic in the number of qubits, while the nonadaptive protocol may use an arbitrary global POVM on each copy and still requires exponentially many copies. The constants are not optimized.

\subsection{A nonadaptive upper bound from 2-design measurements}
\label{sec:ternary_nonadaptive_upper}

To complement the lower bound \(1/B_L\) in~\eqref{eq:ternary_nonadaptive_constant}, it is useful to know how far a nonadaptive global learner can actually get. A rank one POVM whose elements form a complex projective 2-design saturates the Gill--Massar budget of Remark~\ref{rem:gill_massar} and spreads it uniformly over all Pauli directions. Recall that a finite set of unit vectors \(\{|\psi_y\rangle\}_{y\in\mathcal Y}\) in \(\mathbb C^d\) is a complex projective 2-design if
\begin{equation}
\frac{1}{|\mathcal Y|}\sum_{y\in\mathcal Y}\bigl(|\psi_y\rangle\langle\psi_y|\bigr)^{\otimes2}
=
\frac{\Pi_{\mathrm{sym}}}{\binom{d+1}{2}},
\label{eq:two_design}
\end{equation}
where \(\Pi_{\mathrm{sym}}\) is the projector onto the symmetric subspace of \(\mathbb C^d\otimes\mathbb C^d\). The set of all \(L\)-qubit stabilizer states is an example~\cite{GrossAudenaertEisert2007,KuengGross2015}. For such a set, \(\mathcal F=\{F_y\}_{y\in\mathcal Y}\) with \(F_y=\frac{d}{|\mathcal Y|}|\psi_y\rangle\langle\psi_y|\) is a POVM, and \eqref{eq:two_design} gives, for every traceless Hermitian \(A\),
\begin{equation}
\frac{1}{|\mathcal Y|}\sum_{y}\langle\psi_y|A|\psi_y\rangle=0,
\qquad
\frac{1}{|\mathcal Y|}\sum_{y}\langle\psi_y|A|\psi_y\rangle^2=\frac{\Tr(A^2)}{d(d+1)}.
\label{eq:two_design_moments}
\end{equation}

The proof of the benchmark below is given in Appendix~\ref{app:ternary_finite_sample}.

\begin{proposition}[Nonadaptive 2-design benchmark]
\label{prop:ternary_nonadaptive_upper}
Let \(\mathcal F\) be a rank one 2-design POVM on the signal register, and consider the nonadaptive protocol that, on every copy, measures the marker qubits in the \(Z\) basis and the signal register with \(\mathcal F\). Then for every label \(\mathbf a\) and every \(|c|<1\),
\[
I_{\mathcal F}(\mathbf a;c)\geq\frac{1}{(2^L+1)(1+|c|)},
\qquad
I_{\mathcal F}(\mathbf a;0)=\frac{1}{2^L+1},
\]
and there is an estimator for which
\[
\limsup_{N\to\infty}NR^{(L)}_{\mathrm{NA,Glob}}(N;0)\leq2^L+1.
\]
Consequently,
\[
\frac{1}{B_L}
\leq
\liminf_{N\to\infty}NR^{(L)}_{\mathrm{NA,Glob}}(N;0)
\leq
\limsup_{N\to\infty}NR^{(L)}_{\mathrm{NA,Glob}}(N;0)
\leq
2^L+1.
\]
\end{proposition}

Proposition~\ref{prop:ternary_nonadaptive_upper} has two consequences. First, the ratio of nonadaptive to adaptive asymptotic MSE constants for the ternary family is \(2^{\Theta(n)}\): it is at least \((1-2^{-L})^{-1}[3/(2(1+u))]^L\) and at most \(2^L+1\). Second, the same fixed 2-design measurement provides Fisher information \(1/(2^L+1)\) for every label, exponentially more than the \(3^{-L}\) upper bound on worst label Fisher information in the corresponding restricted model of Proposition~\ref{prop:classical_restricted}, where one basis is chosen for each copy. For \(L\geq2\), the quantum nonadaptive learner is therefore stronger than this restricted classical analogue in worst case Fisher information, and the Gill--Massar budget of Remark~\ref{rem:gill_massar} quantifies this advantage. In finite sample terms, conditional on correct marker localization, the estimator has mean square error at most \(2(2^L+1)/N\); combining Chebyshev's inequality with the localization bound shows that \(O(u^{-2}\log L+2^L\epsilon^{-2})\) copies suffice nonadaptively for \eqref{eq:ternary_finite_accuracy}. Together with Theorem~\ref{thm:ternary_finite_sample}, this places the nonadaptive sample complexity at fixed \(u\) and \(\epsilon\) between \(B_L^{-1}/(144\epsilon^2)\) and \(O(2^L\epsilon^{-2})\), both exponential in \(n\).

\section{Structural Properties}
\label{sec:structural_properties}

The preceding sections establish the adaptive/nonadaptive separation by analyzing how measurement information is distributed across a hidden family of signal observables. We now record two structural properties of the state families. These properties are useful because they rule out two possible explanations for the separation: the result does not require entanglement, or any other nonclassical correlation, in the unknown state, and the strongest construction remains compactly representable as a low bond dimension matrix product operator. Along the way we identify what is nonclassical about the construction: not any individual state, but the family of measurements that the hidden label selects. Proofs of the formal results in this section are collected in Appendix~\ref{app:structural_results}.

\subsection{Purely classical state families}
\label{sec:full_separability}

We first note that the signal states used in both constructions are fully separable, and in fact classical: each is diagonal in a product of local orthonormal bases.

\begin{lemma}[Full separability and classicality of single Pauli signal states]
\label{lem:full_separability}
Fix \(L\geq1\), let \( P_{\mathbf w}
=
\bigotimes_{\ell=1}^L w_\ell \), where \( w_\ell\in\{X,Y,Z\} \), and define \( \sigma_{\mathbf w,c}
=
2^{-L}
\left(
\Id_{L}+cP_{\mathbf w}
\right) \), for \( |c|\leq1 \). Then \(\sigma_{\mathbf w,c}\) is fully separable across the \(L\) signal qubits. More precisely, it is diagonal, with nonnegative weights, in the product basis formed by the \(w_\ell\) eigenbases of the individual qubits; it is therefore a strictly classically correlated state, with zero quantum discord across any bipartition of the signal qubits.
\end{lemma}

\begin{corollary}[Classicality of the binary and ternary families]
\label{cor:family_separability}
Every state in the binary family of Section~\ref{sec:binary_state_family} and every state in the ternary family of Section~\ref{sec:ternary_family} is fully separable across all \(2L\) marker and signal qubits, and is diagonal in the label dependent local product basis consisting of the \(Z\) eigenbasis on each marker qubit and the \(w_\ell\) eigenbasis on signal qubit \(\ell\); in particular it carries no quantum discord.
\end{corollary}

Full separability should not be confused with a product structure across the signal qubits. Lemma~\ref{lem:full_separability} shows that \(\sigma_{\mathbf w,c}\) is a classical mixture of product eigenstates, and such a mixture can still contain correlations among the local outcomes. In the present family, the parameter \(c\) controls a correlation that involves all \(L\) signal qubits.

\begin{proposition}[Signal correlation structure]
\label{prop:signal_correlation_structure}
For every \(\mathbf w\in\{X,Y,Z\}^L\) and \(c\in[-1,1]\),
\begin{equation}
\Tr\!\left(P_{\mathbf w}\sigma_{\mathbf w,c}\right)
=
c.
\label{eq:signal_full_correlation}
\end{equation}
Moreover, for any proper subset
\(
T\subsetneq\{1,\ldots,L\}
\)
of the signal sites,
\begin{equation}
\Tr_{T^c}\!\left(\sigma_{\mathbf w,c}\right)
=
\frac{\Id_{|T|}}{2^{|T|}},
\label{eq:signal_proper_marginal}
\end{equation}
independently of both \(c\) and \(\mathbf w\). Consequently, when \(L\geq2\) and \(c\neq0\), the state \(\sigma_{\mathbf w,c}\) is fully separable but is not a product state across the signal qubits.
\end{proposition}

Proposition~\ref{prop:signal_correlation_structure} makes the correlation structure explicit. The parameter \(c\) does not affect any proper marginal of the signal register, but it is exactly the expectation of the full \(L\)-qubit Pauli observable \(P_{\mathbf w}\). Equivalently, if the signal qubits are measured in the local eigenbases of \(w_1,\ldots,w_L\), then \(c\) controls the parity \(\prod_{\ell=1}^L s_\ell\) of the \(L\) outcomes, while every marginal distribution involving fewer than \(L\) outcomes is uniform. Because the marker states in both constructions are independent of \(c\), any reduction of the full state that omits at least one signal qubit is likewise independent of \(c\).

If \(\mathbf w\) were known, this correlation would be straightforward to measure: one could measure signal qubit \(\ell\) in the \(w_\ell\) eigenbasis and multiply the resulting outcomes. The estimation difficulty instead arises because the hidden label determines which observable \(P_{\mathbf w}\) carries the information about \(c\). In the ternary family, the candidate signal observables do not share a common eigenbasis, so a single fixed measurement cannot use the common basis strategy available for the binary family in Section~\ref{sec:binary_global_failure}.

At \(c=0\), the signal state itself is maximally mixed,
\(
\sigma_{\mathbf w,0}=\Id_{L}/2^L
\).
The local parameter \(c\) moves the state away from this center precisely along the \(L\)-body direction
\(
P_{\mathbf w}/2^L
\).
Thus, although the signal state is uncorrelated at \(c=0\), small changes in \(c\) introduce an \(L\)-body correlation that the local asymptotic estimation problem is designed to detect.

Thus state entanglement is not required for any of the adaptive/nonadaptive separations proved above. In particular, the global separation in Theorem~\ref{thm:ternary_separation} occurs even though every unknown state in the ternary family is fully separable and, by Corollary~\ref{cor:family_separability}, even classical. The difficulty lies in how a fixed measurement allocates information across the possible signal observables associated with the hidden labels, not in entanglement of the state itself.

\subsubsection{Classical states and a noncommuting family}
\label{sec:classicality}
Every unknown state is, up to a \emph{label dependent} product unitary, a classical probability distribution: \(\rho_{\mathbf a,c}=U_{\mathbf a}D_{\mathbf a,c}U_{\mathbf a}^\dagger\), where \(D_{\mathbf a,c}\) is diagonal in the computational basis and \(U_{\mathbf a}=\Id_{L}\otimes\bigotimes_{\ell=1}^L V_{a_\ell}\), with \(V_X,V_Y,V_Z\) fixed single qubit Clifford unitaries mapping \(Z\) to \(X\), \(Y\), and \(Z\), respectively. Thus each state is individually classical, but the basis depends on the hidden label. For \(c\neq0\), no single basis diagonalizes all \(3^L\) signal states: if \(\mathbf a\) and \(\mathbf b\) differ at exactly one site, then \(P_{\mathbf a}\) and \(P_{\mathbf b}\) anticommute, so \(\sigma_{\mathbf a,c}\) and \(\sigma_{\mathbf b,c}\) do not commute. This distinction also has a standard operational consequence. By the no-broadcasting theorem~\cite{BarnumCavesFuchsJozsaSchumacher1996}, a family of states can be broadcast by a single channel if and only if its members commute. The common eigenbasis strategy of Section~\ref{sec:binary_global_failure} is the commuting case, whereas the ternary family is noncommuting; Theorem~\ref{thm:ternary_joint_budget} quantifies the resulting limitation on a fixed measurement. The nonclassicality at the level of the family can also be witnessed by discord. For \(c\neq0\), the label flagged ensemble state \(3^{-L}\sum_{\mathbf a}|\mathbf a\rangle\langle\mathbf a|\otimes\sigma_{\mathbf a,c}\) has positive discord with respect to measurements on the signal register because the conditional signal states are not jointly commuting.

\subsubsection{No adaptive advantage for commuting families}
Commutativity is enough to rule out an adaptive advantage under arbitrary single copy measurements.

\begin{proposition}[Exact simulation of adaptive protocols for commuting families]
\label{prop:commuting_simulation}
Let \(\{\omega_\theta:\theta\in\Theta\}\) be a family of pairwise commuting density operators on a finite dimensional Hilbert space; for instance, classical distributions \(p_\theta\) on a sample space \(\Omega\), embedded as the diagonal states \(\sum_{x\in\Omega}p_\theta(x)|x\rangle\langle x|\). For every adaptive protocol using \(N\) independently prepared copies and arbitrary single copy POVMs, there is a nonadaptive protocol that measures all \(N\) copies in a common eigenbasis and uses classical postprocessing to reproduce the adaptive transcript distribution for every \(\theta\). Consequently, adaptive and nonadaptive protocols have identical optimal risk for every estimation target, every loss function, and every \(N\).
\end{proposition}

Two consequences are useful here. First, for classical distributions, observing each sample in full reveals all of the information available from that sample, after which any adaptive protocol can be simulated in post processing. The ternary separation of Theorem~\ref{thm:ternary_separation} therefore cannot be reproduced by a family of classical distributions when the nonadaptive learner observes each sample in full; Fawzi et al.\ make a similar observation for hypothesis selection~\cite{FawziFlammarionGarivierOufkir2023}. Second, the binary family is itself commuting: the \(P_S\) commute for fixed \(|S|=K\), and the marker states are diagonal in the \(Z\) basis. Proposition~\ref{prop:commuting_simulation} therefore gives an exact finite \(N\) counterpart to Theorem~\ref{thm:binary_global_no_separation}, with the common eigenbasis in Section~\ref{sec:binary_global_failure} providing the corresponding nonadaptive measurement. The ternary family is noncommuting, so this exact simulation no longer applies; Theorem~\ref{thm:ternary_joint_budget} turns the absence of a common basis into a quantitative Fisher information limitation. Under restricted measurements, however, a classical analogue remains.

\subsubsection{Classical analogue under restricted measurements}

The restricted product measurement separation can be reproduced by a
classical model with the same locate-then-measure structure.

\begin{proposition}[Classical analogue of the product measurement separation]
\label{prop:classical_restricted}
Consider the following classical model with hidden label \(\mathbf a\in\{x,y\}^L\), parameter \(c\in\Theta=[-c_{\max},c_{\max}]\), and reference point \(c_0\in(-c_{\max},c_{\max})\). A sample consists of marker bits \(m_\ell\in\{\pm1\}\), with mean \(+u\) if \(a_\ell=x\) and \(-u\) if \(a_\ell=y\), and of \(L\) cells \((x_\ell,y_\ell)\in\{\pm1\}^2\). In cell \(\ell\), the \emph{live} bit \(b_\ell\) is the \(a_\ell\) component and the other component is a decoy. The live bits \((b_1,\ldots,b_L)\) have joint law \(2^{-L}\bigl(1+c\prod_\ell b_\ell\bigr)\); the decoy bits are i.i.d.\ uniform and independent of everything else; and the marker bits are independent of the cells. A measurement reads all marker bits and exactly one component of each cell, according to a pattern \(\mathbf r\in\{x,y\}^L\) that must be fixed before the sample is seen. Here \(R_{\mathrm{AD}}\) and \(R_{\mathrm{NA}}\) denote the local minimax MSEs of \eqref{eq:generic_local_minimax_mse} for this restricted model, under adaptive and nonadaptive measurement patterns, respectively. Then the Fisher information of one sample about \(c\) equals \(1/(1-c^2)\) if \(\mathbf r=\mathbf a\) and \(0\) otherwise; an adaptive protocol achieves \(\lim_{N\to\infty}NR_{\mathrm{AD}}(N;c_0)=1-c_0^2\), which no protocol can improve; and every nonadaptive protocol satisfies \(\liminf_{N\to\infty}NR_{\mathrm{NA}}(N;c_0)\geq2^L(1-c_0^2)\). An analogous model with cells \((x_\ell,y_\ell,z_\ell)\), hidden label \(\mathbf a\in\{x,y,z\}^L\), marker means \(-u,0,+u\) for the three local labels, and exactly one observed component per cell gives the same statements with \(2^L\) replaced by \(3^L\).
\end{proposition}

The model in Proposition~\ref{prop:classical_restricted} is the classical distribution obtained by writing the signal state in its local eigenbases, where the bias appears only in the parity, and reading one component from each cell is the classical counterpart of choosing one local basis at each signal site. Thus the restricted separation does not by itself require a quantum effect; the restriction on what is observed is enough to produce it. Allowing the classical learner to read the markers of a sample before choosing that sample's pattern can increase its aggregate budget by at most \((1+u)^L\), the factor in Lemma~\ref{lem:ternary_marker_factor}; this is the classical part of the decomposition in Remark~\ref{rem:budget_decomposition}. A general quantum POVM is not restricted to one candidate basis per copy and can carry information about several candidate observables, while the Gill--Massar budget limits the aggregate Fisher information available across them.

\subsubsection{Measurement incompatibility}
These results identify measurement incompatibility as the quantum ingredient behind the global separation. The adaptive learner can use marker outcomes from earlier copies to identify the hidden label before choosing later signal measurements. A nonadaptive learner may obtain the same marker information, but its single-copy POVMs are fixed before any outcomes are observed, so the label can affect only the final classical post processing.

In state discrimination, the analogous distinction is between pre-measurement and post-measurement information~\cite{BallesterWehnerWinter2008,GopalWehner2010}. Carmeli, Heinosaari, and Toigo~\cite{CarmeliHeinosaariToigo2018,CarmeliHeinosaariToigo2019} show that, in the settings they study, the two scenarios have the same optimal performance if and only if the optimal measurements for the different ensembles are compatible, or jointly measurable; a gap therefore witnesses incompatibility. The binary family is the compatible case because the \(P_S\) commute and share a common eigenbasis, whereas the ternary construction uses incompatible local Pauli measurements. Theorem~\ref{thm:ternary_separation} gives the corresponding estimation-theoretic manifestation: an exponential adaptive/nonadaptive local asymptotic MSE gap whose nonadaptive lower bound is controlled by the aggregate Fisher-information budget of Theorem~\ref{thm:ternary_joint_budget}. Measurement incompatibility is distinct from entanglement, nonclassical correlations in the state, and quantum memory across copies~\cite{HeinosaariMiyaderaZiman2016,SkrzypczykSupicCavalcanti2019,BuscemiChitambarZhou2020}; among these resources, it is the one used by the construction.

\subsection{Exact MPO structure}
\label{sec:ternary_mpo_structure}

The ternary family is also compactly representable as a matrix product operator (MPO)~\cite{VerstraeteGarciaRipollCirac2004,QinJamesonGongWakinZhu2024}. Group marker qubit \(\ell\) and signal qubit \(\ell\) into one local block, so that the \(L\) blocks each have local Hilbert space dimension four. After reordering the tensor factors into these paired blocks, define \( A_\ell
:=
\tau_{a_\ell}\otimes\Id_{1} \), and \( B_\ell
:=
\tau_{a_\ell}\otimes a_\ell \), where \(\Id_{1}\) in this case denotes the single qubit identity on signal site \(\ell\). Equation~\eqref{eq:ternary_state_family} then becomes
\begin{equation}
\rho_{\mathbf a,c}
=
2^{-L}
\left[
\bigotimes_{\ell=1}^L A_\ell
+
c
\bigotimes_{\ell=1}^L B_\ell
\right].
\label{eq:ternary_two_product_mpo}
\end{equation}

\begin{proposition}[Exact MPO Representation with Bond Dimension Two]
\label{prop:ternary_mpo}
For every \(\mathbf a\in\{X,Y,Z\}^L\) and every \(c\in[-1,1]\), the state \(\rho_{\mathbf a,c}\), under the marker--signal blocking above, admits an exact MPO representation as an operator with bond dimension \(\chi_{\mathrm{MPO}}\leq2\). \end{proposition}

We therefore see that adaptive/nonadaptive separation can occur inside a family whose density operators have an exact constant bond dimension representation under a natural local blocking. We comment further on this connection in Section~\ref{sec:mpometrics}.

\section{State Reconstruction Metrics and Statistical Conditioning}
\label{sec:reconstruction_metrics}

The separations we have established are stated in terms of the local asymptotic minimax MSE for the scalar parameter \(c\), whereas much of the structured tomography literature measures recovery error directly in a metric on the reconstructed density matrix. It is useful to make the relation between these two notions of accuracy explicit. The key point is that a compact (e.g., MPO) representation does not necessarily make every parameter in the state easy to estimate.

\subsection{Exact metric relations for the ternary family}

Recall that \(d=2^L\), that \(u\) is the fixed marker strength parameter from the ternary construction with \(0<u<\tfrac12\), and that \( \rho_{\mathbf a,c}
=
\tau_{\mathbf a}\otimes \frac{\Id_{L}+cP_{\mathbf a}}{d} \). For a word \(\mathbf a\in\{X,Y,Z\}^L\), define \( m(\mathbf a)
:=
\big|\{\ell:a_\ell\in\{X,Z\}\}\big| \). The following identities quantify exactly how a change in \(c\) appears in common state space metrics. Their proof is given in Appendix~\ref{app:metric_relations}.

\begin{proposition}[Metric geometry of the ternary family]
\label{prop:ternary_metric_geometry}
For every fixed \(\mathbf a\in\{X,Y,Z\}^L\) and \(c,c'\in[-1,1]\),
\begin{align}
\frac12\left\|\rho_{\mathbf a,c}-\rho_{\mathbf a,c'}\right\|_1
&=
\frac{|c-c'|}{2},
\label{eq:ternary_trace_metric_relation}
\\
\left\|\rho_{\mathbf a,c}-\rho_{\mathbf a,c'}\right\|_F^2
&=
\frac{(c-c')^2(1+u^2)^{m(\mathbf a)}}{4^L}.
\label{eq:ternary_frob_metric_relation}
\end{align}
Moreover,
\begin{equation}
\left\|\rho_{\mathbf a,c}\right\|_F^2
=
\frac{(1+u^2)^{m(\mathbf a)}(1+c^2)}{4^L},
\label{eq:ternary_state_frob_norm}
\end{equation}
and hence, when two states are compared within the same fixed label family,
\begin{equation}
\frac{\left\|\rho_{\mathbf a,c}-\rho_{\mathbf a,c'}\right\|_F^2}
{\left\|\rho_{\mathbf a,c}\right\|_F^2}
=
\frac{(c-c')^2}{1+c^2}.
\label{eq:ternary_relative_frob_relation}
\end{equation}
\end{proposition}

The trace distance relation is dimension independent: changing \(c\) by an \(O(1)\) amount changes the state by an \(O(1)\) trace distance. In contrast, the absolute Frobenius distance suppresses the same change exponentially in \(L\). This distinction is important when comparing the present estimation problem with reconstruction guarantees stated in different metrics.

\subsection{From state reconstruction to estimation of \(c\)}

The metric relations above can also be expressed directly as stability bounds for recovering \(c\) from an estimated state. First suppose that the hidden word \(\mathbf a\) is known. For any estimate \(\widehat\rho\), define \( \widehat c_{1}
:=
\Tr\!\left[(\Id_{L} \otimes P_{\mathbf a})\widehat\rho\right] \). Since \( \Tr\!\left[(\Id_{L} \otimes P_{\mathbf a})\rho_{\mathbf a,c}\right]=c \) and \(\|\Id_{L} \otimes P_{\mathbf a}\|_\infty=1\), using the standard trace norm bound,
\begin{equation}
|\widehat c_1-c|
\le
\left\|\widehat\rho-\rho_{\mathbf a,c}\right\|_1.
\label{eq:c_from_trace_reconstruction}
\end{equation}
Thus trace norm reconstruction accuracy transfers to the accuracy of \(c\) without any loss that depends on the dimension.

For Frobenius error, it is more natural to project onto the actual tangent direction of the fixed label family. Define \( \widehat c_F
:=
\frac{\Tr\!\left[(\tau_{\mathbf a}\otimes P_{\mathbf a})\widehat\rho\right]}
{\Tr(\tau_{\mathbf a}^2)} \). This estimator is exact on the model family, and Cauchy-Schwarz gives
\begin{equation}
|\widehat c_F-c|
\le
\frac{d}{(1+u^2)^{m(\mathbf a)/2}}
\left\|\widehat\rho-\rho_{\mathbf a,c}\right\|_F.
\label{eq:c_from_frob_reconstruction}
\end{equation}
Since \(m(\mathbf a)\le L\),
\begin{equation}
\frac{d}{(1+u^2)^{m(\mathbf a)/2}}
\ge
\left(\frac{2}{\sqrt{1+u^2}}\right)^L,
\label{eq:frob_conditioning_exponential}
\end{equation}
which grows exponentially for the fixed range \(0<u<\tfrac12\). Consequently, a reconstruction guarantee that is polynomial in system size for a fixed \emph{absolute} Frobenius target does not, by itself, imply polynomially well conditioned estimation of \(c\). To force an \(O(1)\) error in \(c\) through \eqref{eq:c_from_frob_reconstruction}, the required absolute Frobenius accuracy is exponentially small in \(L\).

This conditioning comparison does not change the order of limits in the main separation theorem. As throughout the paper, \(L\) and \(u\) are fixed before \(N\to\infty\); the exponential statement concerns how the resulting local asymptotic leading constant scales with \(L\), rather than a joint finite sample limit in which \(N\) and \(L\) grow together.

Equation~\eqref{eq:ternary_relative_frob_relation} also shows why absolute and relative Frobenius criteria should be distinguished. Within a fixed label model family, relative Frobenius accuracy tracks the \(c\)-error on an \(O(1)\) scale. Indeed, combining \eqref{eq:c_from_frob_reconstruction} with \eqref{eq:ternary_state_frob_norm} shows that
\[
\left\|\widehat\rho-\rho_{\mathbf a,c}\right\|_F
\le
\varepsilon\left\|\rho_{\mathbf a,c}\right\|_F
\quad\Longrightarrow\quad
|\widehat c_F-c|
\le
\varepsilon\sqrt{1+c^2},
\]
provided the correct word \(\mathbf a\) is used in the projection. Thus a uniform relative Frobenius guarantee would be substantially stronger for the present task than a fixed absolute Frobenius guarantee.

Although the word \(\mathbf a\) is unknown, a sufficiently accurate trace norm reconstruction is enough to recover it from the marker register. Let \(Z_\ell^{\mathrm{mark}}\) denote \(Z\) on marker site \(\ell\) and the identity elsewhere. Under the three marker states,
\[
\Tr(Z\tau_X)=-u,\qquad
\Tr(Z\tau_Y)=0,\qquad
\Tr(Z\tau_Z)=u.
\]
Hence, if \( \left\|\widehat\rho-\rho_{\mathbf a,c}\right\|_1<\frac{u}{2} \), then every marker expectation \(\Tr(Z_\ell^{\mathrm{mark}}\widehat\rho)\) lies within \(u/2\) of its true value, so nearest neighbor decoding among \(\{-u,0,u\}\) recovers the entire word \(\mathbf a\). A uniform trace norm reconstruction guarantee with favorable system size scaling would therefore be directly relevant to the present estimation problem.

\subsection{Relation to MPO reconstruction results}
\label{sec:mpometrics}

The metric relations above clarify the relation to existing structured tomography results. Qin et al.~\cite{QinJamesonGongWakinZhu2024} established polynomial copy recovery guarantees for bounded \emph{absolute} Frobenius error of constant bond dimension MPOs using Haar random global projective measurements. However, their trace norm extension relies on low matrix rank and is explicitly noted to become vacuous for high rank states. More recent sample efficient MPO tomography results using global informationally complete POVMs and spherical \(t\)-designs likewise give uniform guarantees in absolute Frobenius norm, with trace distance guarantees obtained under an additional approximate low rank assumption~\cite{QinJamesonGoldarWakinGongZhu2024}. These measurement schemes are fully compatible with the global single copy measurement class considered here; the distinction lies in the reconstruction metric and in how accurately that metric controls the estimation error in \(c\).

For the present family, this low matrix rank route does not apply directly. For \(|c|<1\), which contains the local asymptotic neighborhood for all sufficiently large \(N\), the signal state has rank \(d\), and because \(0<u<\tfrac12\), the marker state also has rank \(d\). Hence
\begin{equation}
\operatorname{rank}(\rho_{\mathbf a,c})=d^2=4^L.
\label{eq:ternary_full_matrix_rank}
\end{equation}
Thus the constant MPO bond dimension established in Section~\ref{sec:structural_properties} coexists with full matrix rank. Low MPO bond dimension controls representation complexity; it does not imply low matrix rank or favorable conditioning in every state space metric.

Local reduction tomography provides a complementary example. Baumgratz et al.~\cite{BaumgratzGrossCramerPlenio2013} showed that broad classes of MPOs satisfying an invertibility condition can be reconstructed from fixed size contiguous local reductions. By Proposition~\ref{prop:signal_correlation_structure} and the fact that the marker state is independent of \(c\), any reduction of the present family that omits even one signal qubit is independent of \(c\). Thus the local reduction mechanism considered there cannot identify \(c\) from such reductions.

These observations also distinguish informational completeness from finite sample estimation accuracy. An informationally complete POVM uniquely determines the state from its exact outcome probabilities, but this one-to-one property alone does not guarantee strong statistical sensitivity in every structured direction.

The structured tomography results above and the separation proved here are therefore complementary. Existing polynomial copy complexity guarantees for MPOs establish efficient reconstruction under particular metrics and structural assumptions, whereas our result concerns the local minimax MSE for the scalar parameter \(c\), whose informative direction depends on the hidden word. For a reconstruction result to be directly comparable with our asymptotic bound, its error metric would need to control the recovery of both the hidden word and \(c\) with accuracy that does not deteriorate unfavorably with the system size, under the same nonadaptive single copy measurement model. The exact relations above show why a fixed absolute Frobenius recovery guarantee alone does not provide such control for this family.

\section{Discussion and Conclusions}
\label{sec:discussion_conclusions}

\subsection{Root of the adaptive advantage}
\label{sec:discussion_mechanism}

The binary and ternary constructions together clarify the mechanism behind the adaptive/nonadaptive separation. In both families, the marker register carries information about a hidden label, while the signal register carries information about \(c\) through a signal observable determined by that label. For fixed structural parameters, the hidden label can be localized from the marker register using only \(O(\log N)\) copies. An adaptive protocol can therefore spend a vanishing fraction of the sample budget learning which signal observable is relevant and then use all remaining copies to measure that observable.

The measurement hierarchy shows why hiding the informative observable is not, by itself, sufficient. Under product measurements, the binary family already creates a bottleneck in information allocation. Once arbitrary global single copy measurements are allowed, however, the commuting binary signal observables admit a common eigenbasis. A fixed measurement can then record the candidate eigenvalue statistics simultaneously and postpone identification of the relevant label to classical post processing. The ternary construction removes this opportunity because its candidate signal observables do not share a common eigenbasis. This incompatibility explains why the binary global strategy no longer applies, but it does not by itself yield the nonadaptive lower bound. The required quantitative statement comes from Theorem~\ref{thm:ternary_joint_budget}: under any fixed global single copy POVM, the \(3^L\) distinct Hilbert--Schmidt orthogonal Pauli observables share a limited aggregate Fisher information budget, even when the marker register is included.

Averaging this budget forces at least one hidden label to receive exponentially small Fisher information, whereas adaptation first learns the label and then concentrates later measurements on the corresponding signal observable. Section~\ref{sec:structural_properties} shows that this effect occurs even though the states are fully separable and the ternary family admits an exact MPO representation with bond dimension at most two under the natural blocking. Crucially, full separability does not imply the absence of correlations: Proposition~\ref{prop:signal_correlation_structure} shows that \(c\) is the expectation of the full \(L\)-qubit signal observable, while every proper signal marginal is independent of \(c\). The separation is therefore driven by hidden measurement structure and the ability to reallocate future measurement effort after that structure has been learned, rather than by entanglement in the unknown state.

Section~\ref{sec:full_separability}  sharpens this conclusion. The states are not merely separable but classical: each is diagonal in a label dependent local product basis, so it carries no quantum discord or any other nonclassical correlation. No commuting family, classical or quantum, admits an adaptive advantage under unrestricted single copy measurements (Proposition~\ref{prop:commuting_simulation}), so the global separation has no classical analogue, whereas the product measurement separation does (Proposition~\ref{prop:classical_restricted}). What remains, and what the ternary family uses, is the incompatibility of the family of signal measurements indexed by the label: the adaptive and nonadaptive learners are the pre- and post-measurement-information receivers of~\cite{CarmeliHeinosaariToigo2018}, and the aggregate budget is the Gill--Massar inequality~\cite{GillMassar2000}. In the decomposition \(3^LB_L=(2^L-1)(1+u)^L\) of Remark~\ref{rem:budget_decomposition}, the factor \((1+u)^L\) is classical and the factor \(2^L-1\) is the quantum budget. Proposition~\ref{prop:ternary_nonadaptive_upper} shows that this quantum budget is real: a 2-design nonadaptive learner reaches Fisher information \(1/(2^L+1)\) for every label, exponentially more than any learner committed to one candidate basis per copy, yet still exponentially less than the adaptive learner. The resource behind the separation is thus measurement incompatibility, not any correlation property of the unknown state.

\subsection{Scope and broader implications}
\label{sec:discussion_scope_implications}

The strongest global result is established for the local asymptotic minimax MSE at \(c=0\) with \(0<u<\tfrac12\), with \(L\) and \(u\) fixed while \(N\to\infty\). The measurement model is single copy: arbitrary global POVMs may act on all marker and signal qubits within each copy, while coherent collective measurements across different copies are not part of the model. These conditions define the regime in which the exponential scaling of the asymptotic leading constants is proved. Theorem~\ref{thm:ternary_finite_sample} complements this asymptotic statement with a finite sample one that involves no limit: at fixed marker strength and accuracy, \(O(\log n)\) copies suffice adaptively while \(2^{\Omega(n)}\) are necessary nonadaptively, for every \(L\). The two statements use different performance criteria and neither implies the other, but both rest on the same information budget.

Relative to the Pauli restricted separation of our earlier work~\cite{GoldarQinZhuGongWakin2026}, the main difference is not the measurement class alone but what the lower bound rests on. A restriction to Pauli settings makes it possible to count settings and show that some are rarely used; without the restriction there is nothing to count, and the nonadaptive learner can in principle spread each copy over all candidate observables at once. The Gill--Massar budget (Lemma~\ref{lem:ternary_signal_budget} and Remark~\ref{rem:gill_massar}) is what limits this spreading, and incompatibility of the candidate observables is what makes the budget binding. Since the prefix observables of~\cite{GoldarQinZhuGongWakin2026} are themselves incompatible across hidden strings, we expect, but have not proved, that a global version of that separation also holds. Finally, the signal states used here also appear as the hard instance of the two copy separation in~\cite{ChenGongZhang2024}; there the hidden Pauli string is located by Bell sampling on pairs of copies, so the present construction can be viewed as a single copy counterpart in which a classical marker performs that step, with one round of adaptation and a trivially computable adaptive rule.

Within this regime, the result shows that a compact state representation does not eliminate a difficult measurement allocation problem. The ternary family is fully separable and has an exact constant bond dimension MPO representation, yet the signal observable that carries information about \(c\) depends on hidden structure that must first be learned. At \(c=0\), the signal state itself is maximally mixed and uncorrelated, but small changes in \(c\) introduce the \(L\)-body correlation identified in Proposition~\ref{prop:signal_correlation_structure}. This isolates a mechanism that can be relevant more broadly in structured quantum estimation: early observations can identify which measurements should receive the remaining experimental effort. The present construction makes that mechanism explicit enough to prove an exponential adaptive/nonadaptive separation.

Section~\ref{sec:reconstruction_metrics} also clarifies the distinction between this scalar estimation problem and full state reconstruction under different error metrics. Sufficiently accurate trace norm reconstruction can recover the hidden word and control the error in \(c\) without a dimension dependent loss, whereas controlling the absolute Frobenius error to a fixed tolerance does not, by itself, control the error in \(c\) with dimension independent accuracy. Thus the polynomial copy reconstruction guarantees in absolute Frobenius norm discussed there and the present local asymptotic MSE separation concern different accuracy criteria and are compatible. In this sense, the result isolates a structured scalar estimation problem within the broader setting of quantum tomography in which measurement allocation, rather than representation complexity alone, is decisive.

\subsection{Open directions}
\label{sec:discussion_open_directions}

A natural next question is whether the strongest global separation can be extended to local neighborhoods centered away from \(c=0\), or to larger marker strengths through a sharper analysis or a different construction. It is also important to understand what changes when coherent collective measurements across multiple copies are permitted.

More broadly, an important question is whether the same information allocation mechanism appears in more general structured state models, where the relevant measurement is not indexed by an explicit finite hidden label. General MPO families provide a natural setting in which to study this question. A separate challenge is to understand the mechanism at finite sample sizes, where the relevant structure may be identified with nonnegligible error. It would also be useful to characterize robustness to model mismatch and regimes in which the system size grows with the number of copies. Addressing these questions would help connect the present information theoretic separation to practical adaptive tomography.

Several further directions are suggested by Sections~\ref{sec:ternary_finite_sample} and~\ref{sec:full_separability}. First, Theorem~\ref{thm:ternary_finite_sample} is stated for estimation of \(c\); a trace distance identification version in the format of~\cite{GoldarQinZhuGongWakin2026}, using the metric relations of Section~\ref{sec:reconstruction_metrics}, would make the two papers directly comparable. Second, the converse question is whether the prefix/tree family of~\cite{GoldarQinZhuGongWakin2026} remains exponentially separated when the nonadaptive learner may use arbitrary global single copy POVMs; its candidate observables are incompatible across hidden strings, and the Gill--Massar budget suggests a base \(2\) rather than \(3\) per qubit, but the in band breadcrumbs complicate the budget accounting and the question is open. Third, the gap between the lower bound \(1/B_L\) and the 2-design upper bound \(2^L+1\) of Proposition~\ref{prop:ternary_nonadaptive_upper} remains; closing it would determine the exact base of the exponential. Fourth, since the separation is an incompatibility witness, it would be interesting to relate its size to quantitative measures of incompatibility of the family of candidate measurements, such as incompatibility robustness~\cite{HeinosaariMiyaderaZiman2016}, and to ask whether every sufficiently incompatible and cheaply indexed family of measurements yields an exponential adaptive advantage. Finally, the present construction needs only one round of adaptation; whether additional rounds can be strictly more powerful in a single copy setting is open.

The main conclusion is that simplicity of the state class does not imply simplicity of the measurement problem. Even for fully separable, indeed classical, states with a constant bond dimension MPO representation, a scalar parameter can be encoded in an \(L\)-body correlation whose informative observable depends on hidden structure revealed only through the data. In the family studied here, adaptation uses early measurements to identify where later measurement effort should be concentrated, yielding an exponential separation in the leading constant of the local asymptotic minimax MSE. The separation is a genuinely quantum phenomenon, but its quantumness resides in the incompatibility of the measurements indexed by the hidden label rather than in the unknown states themselves.

\appendices

\section{Fisher Information and Local Minimax MSE}
\label{app:fisher_mse}

The following lemma converts a Fisher information upper bound into the local asymptotic minimax MSE lower bound used in the separation results.

\begin{lemma}[Local minimax bound from a Fisher information budget]
\label{lem:local_fisher_to_mse}
Fix \(c_0\) in the interior of the parameter space and let
\[
w_N=N^{-1/2+\epsilon},
\qquad
0<\epsilon<\frac12.
\]
Assume that the outcome distributions are differentiable in \(c\) on the local window and have finite Fisher information, so that the van Trees inequality applies. Suppose that, uniformly over the local window,
\begin{equation}
\sup_{|c-c_0|\leq w_N}
J_N(c)
\leq
N\bigl(A+o(1)\bigr)
\label{eq:fisher_budget_generic}
\end{equation}
for some constant \(A>0\), where \(J_N(c)\) denotes the total Fisher information about \(c\) contained in the \(N\) measurement outcomes. Then every estimator satisfies
\begin{equation}
\liminf_{N\to\infty}
N
\sup_{|c-c_0|\leq w_N}
\E_c
\left[
(\widehat c_N-c)^2
\right]
\geq
\frac{1}{A}.
\label{eq:fisher_to_mse}
\end{equation}
No unbiasedness assumption is required.
\end{lemma}

\begin{proof}
Choose a fixed smooth probability density supported on \([-1,1]\) that vanishes smoothly at the endpoints, and rescale and shift it so that its support becomes \( [c_0-w_N,c_0+w_N] \). Denote the resulting density by \(\pi_N\). This transformation preserves the shape of the density while compressing its width by a factor \(w_N\). Consequently, its Fisher information scales as \( I(\pi_N)
=
O(w_N^{-2}) \). Since \(w_N=N^{-1/2+\epsilon}\) with \(\epsilon>0\),
\[
w_N^{-2}
=
N^{1-2\epsilon}
=
o(N),
\]
and therefore \(I(\pi_N)=o(N)\). The van Trees inequality~\cite{VanTrees1968,GillLevit1995} gives
\[
\int
\E_c
\left[
(\widehat c_N-c)^2
\right]
\pi_N(c)\,dc
\geq
\frac{1}{
\int J_N(c)\pi_N(c)\,dc
+
I(\pi_N)
}.
\]
By the assumed Fisher information bound
\eqref{eq:fisher_budget_generic},
\[
\sup_{|c-c_0|\leq w_N}
J_N(c)
\leq
N\bigl(A+o(1)\bigr),
\]
and since \(\pi_N\) is supported entirely on this window,
\[
\int J_N(c)\pi_N(c)\,dc
\leq
N\bigl(A+o(1)\bigr).
\]
Combining this with \(I(\pi_N)=o(N)\) yields
\[
\int
\E_c
\left[
(\widehat c_N-c)^2
\right]
\pi_N(c)\,dc
\geq
\frac{1}{
NA+o(N)
}.
\]

The worst-case MSE over the local window is at least its average under \(\pi_N\). Hence
\[
\sup_{|c-c_0|\leq w_N}
\E_c
\left[
(\widehat c_N-c)^2
\right]
\geq
\frac{1}{
NA+o(N)
}.
\]
Multiplying by \(N\) and taking the limit inferior as \(N\to\infty\) gives
\[
\liminf_{N\to\infty}
N
\sup_{|c-c_0|\leq w_N}
\E_c
\left[
(\widehat c_N-c)^2
\right]
\geq
\frac{1}{A},
\]
which proves \eqref{eq:fisher_to_mse}.
\end{proof}

\section{Proofs and Supporting Results for the Binary Family}
\label{app:binary_results}

\subsection{Adaptive localization and product Pauli measurements}
\label{app:binary_pauli}

\paragraph{Lemma~\ref{lem:binary_marker_localization}}
\begin{proof}
Because the marker state is a product state across sites and each localization copy is independently prepared, for every fixed site \(\ell\) the random variables
\(Z_{\ell,1},\ldots,Z_{\ell,m_0}\) are independent, take values in \([-1,1]\), and satisfy
\[
\E_S[Z_{\ell,t}]
=
\alpha_{\ell,S}
=
\begin{cases}
u,&\ell\in S,\\
0,&\ell\notin S.
\end{cases}
\]
Hence \(\E_S[\widehat\alpha_\ell]
=
\alpha_{\ell,S}\). Define the good event \( \mathcal G
=
\bigcap_{\ell=1}^{L}
\left\{
\left|
\widehat\alpha_\ell-\alpha_{\ell,S}
\right|
<
\frac{u}{2}
\right\} \). On \(\mathcal G\), every active site \(\ell\in S\) satisfies \( \widehat\alpha_\ell
>
u-\frac{u}{2}
=
\frac{u}{2} \), whereas every inactive site \(\ell\notin S\) satisfies \( \widehat\alpha_\ell
<
0+\frac{u}{2}
=
\frac{u}{2} \). Thus every active empirical marker score is strictly larger than every inactive one. In particular, no tie between an active and an inactive site can occur on \(\mathcal G\), so the choice of tie-breaking rule is irrelevant on this event. Since \(|S|=K\), the \(K\) largest empirical scores are exactly those indexed by \(S\). Therefore \( \mathcal G
\subseteq
\{\widehat S=S\} \), or equivalently, \(\{\widehat S\neq S\}
\subseteq
\mathcal G^c\). 
For each fixed site, Hoeffding's inequality for independent random variables in \([-1,1]\) gives
\[
\Pr_S\!\left(
\left|
\widehat\alpha_\ell-\alpha_{\ell,S}
\right|
\geq
\frac{u}{2}
\right)
\leq
2\exp\!\left(
-\frac{m_0u^2}{8}
\right).
\]
Applying a union bound over the \(L\) marker sites yields
\[
\Pr_S(\widehat S\neq S)
\leq
\Pr_S(\mathcal G^c)
\leq
2L
\exp\!\left(
-\frac{m_0u^2}{8}
\right),
\]
which proves \eqref{eq:binary_marker_error}. The bound does not depend on the identity of \(S\), so it is uniform over \(\mathcal S_{L,K}\).

Now choose \(m_0(N)\) as in \eqref{eq:binary_marker_budget}. Since
\[
m_0(N)
\geq
\frac{8}{u^2}
\left[
\log(2L)+2\log(N+2)
\right],
\]
we have
\[
2L
\exp\!\left(
-\frac{m_0(N)u^2}{8}
\right)
\leq
2L
\exp\!\bigl(
-\log(2L)-2\log(N+2)
\bigr)
=
(N+2)^{-2}.
\]
Hence
\[
\sup_{S\in\mathcal S_{L,K}}
\Pr_S(\widehat S\neq S)
\leq
(N+2)^{-2}
=
o(N^{-1}).
\]

Finally, with \(L\), \(K\), and \(u\) fixed before taking \(N\to\infty\), we obtain \( m_0(N)=O(\log N)=o(N) \). Thus the support can be localized with error probability negligible on the \(N^{-1}\) risk scale while using only a vanishing fraction of the available copies.
\end{proof}

\paragraph{Lemma~\ref{lem:binary_pauli_information}}
\begin{proof}
If the setting matches \(P_S\), the parity statistic has distribution
\[
\Pr(R=+1)=\frac{1+c}{2},
\qquad
\Pr(R=-1)=\frac{1-c}{2},
\]
whose Fisher information for \(c\) is \( \frac{1}{1-c^2} \). 
If even one local Pauli axis differs from the corresponding factor of \(P_S\), the \(cP_S\) term in \(\sigma_{S,c}\) has zero overlap with every outcome probability of that setting. The resulting outcome distribution is therefore independent of \(c\), and its Fisher information is zero.
\end{proof}

\paragraph{Theorem~\ref{thm:binary_pauli_separation}}
\begin{proof}
We first prove adaptive achievability. Use \(m_0\) copies for marker localization as in
Lemma~\ref{lem:binary_marker_localization}, then measure \(P_{\widehat S}\) on the remaining \(N-m_0\) copies and estimate \(c\) from their parity sample mean.

On the event \(\widehat S=S\), clipping the sample mean to \(\Theta\) cannot increase its squared error, and hence
\[
\E_{S,c}
\left[
(\widehat c_N-c)^2
\mid
\widehat S=S
\right]
\leq
\frac{1-c^2}{N-m_0}.
\]
On the event \(\widehat S\neq S\), both \(c\) and the clipped estimate lie in
\([-c_{\max},c_{\max}]\), so \( (\widehat c_N-c)^2
\leq
4c_{\max}^2 \). Therefore, uniformly over \(S\),
\[
\sup_{|c-c_0|\leq w_N}
\E_{S,c}
\left[
(\widehat c_N-c)^2
\right]
\leq
\frac{
\sup_{|c-c_0|\leq w_N}(1-c^2)
}{
N-m_0
}
+
4c_{\max}^2
\Pr_S(\widehat S\neq S).
\]
Since
\[
m_0=o(N),
\qquad
N\Pr_S(\widehat S\neq S)\to0,
\qquad
w_N\to0,
\]
we obtain
\[
\limsup_{N\to\infty}
N
R_{\mathrm{AD,PP}}^{(L,K)}(N;c_0)
\leq
1-c_0^2.
\]

For the matching lower bound, fix any support \(S\). The marker state does not depend on \(c\), so it contributes no quantum Fisher information about \(c\). Moreover, \( \sigma_{S,c}
=
2^{-L}(\Id_{L}+cP_S) \) has a \(c\)-independent eigenbasis because \(P_S\) is fixed and has eigenvalues \(\pm1\). The corresponding eigenvalues of \(\sigma_{S,c}\) are \( \frac{1+c}{2^L} \) and \( \frac{1-c}{2^L} \)
each with multiplicity \(2^{L-1}\). Hence the quantum Fisher information for \(c\) is \(H(c)
=
\frac{1}{1-c^2}\). Therefore no single copy measurement can have classical Fisher information larger than \(H(c)\).

For an adaptive protocol, conditioning on all previous outcomes fixes the POVM applied to the next independently prepared copy. Each conditional measurement therefore contributes at most \(H(c)\) Fisher information, and the conditional contributions add across copies. Thus \( J_N(c)
\leq
\frac{N}{1-c^2} \). Since
\[
\sup_{|c-c_0|\leq w_N}
\frac{1}{1-c^2}
=
\frac{1}{1-c_0^2}
+
o(1),
\]
Lemma~\ref{lem:local_fisher_to_mse} applies with \( A=\frac{1}{1-c_0^2} \). Taking the worst case over \(S\) gives
\[
\liminf_{N\to\infty}
N
R_{\mathrm{AD,PP}}^{(L,K)}(N;c_0)
\geq
1-c_0^2.
\]
Together with the achievability bound, this proves
\eqref{eq:binary_adaptive_constant}.

Now consider an arbitrary nonadaptive product Pauli protocol. For each
\(S\in\mathcal S_{L,K}\), let \(\bar n_S\) denote the expected number of copies whose signal setting matches \(P_S\). This definition also covers a schedule randomized before data collection; the realized measurement choices are part of the observed record, as in Section~\ref{sec:adaptive_nonadaptive}.

Each signal setting can match at most one support, so \( \sum_{S\in\mathcal S_{L,K}}
\bar n_S
\leq
N \). Consequently, for every such protocol there exists a support
\(S_N^\star\in\mathcal S_{L,K}\) satisfying
\begin{equation}
\bar n_{S_N^\star}
\leq
\frac{N}{M_{L,K}}.
\label{eq:binary_worst_support_pauli}
\end{equation}

Marker outcomes may help identify \(S_N^\star\) after the measurements have been collected, but their distribution does not depend on \(c\), so they carry no Fisher information about \(c\). By
Lemma~\ref{lem:binary_pauli_information}, only copies whose signal setting matches \(P_{S_N^\star}\) contribute Fisher information for \(c\). Hence \( J_{S_N^\star}(c)
\leq
\frac{N}{
M_{L,K}(1-c^2)
} \). It follows that, uniformly over the local window,
\[
\sup_{|c-c_0|\leq w_N}
J_{S_N^\star}(c)
\leq
N
\left[
\frac{1}{
M_{L,K}(1-c_0^2)
}
+
o(1)
\right].
\]
Here the \(o(1)\) term depends only on \(c_0\) and the shrinking window \(w_N\), and is therefore uniform over all admissible nonadaptive protocols.

Applying Lemma~\ref{lem:local_fisher_to_mse} with \( A
=
\frac{1}{
M_{L,K}(1-c_0^2)
} \) gives
\[
\liminf_{N\to\infty}
N
\sup_{|c-c_0|\leq w_N}
\E_{S_N^\star,c}
\left[
(\widehat c_N-c)^2
\right]
\geq
M_{L,K}(1-c_0^2).
\]
Because the criterion in \eqref{eq:generic_local_minimax_mse} takes the worst case over all supports, this lower bound applies to every admissible nonadaptive protocol and estimator. The uniformity of the Fisher information bound above allows us to take the infimum over protocols and estimators without changing the asymptotic lower bound. Therefore
\[
\liminf_{N\to\infty}
N
R_{\mathrm{NA,PP}}^{(L,K)}(N;c_0)
\geq
M_{L,K}(1-c_0^2),
\]
which proves
\eqref{eq:binary_nonadaptive_pauli_constant}.
\end{proof}

\subsection{Arbitrary product POVMs}
\label{app:binary_product_povm}

Consider a single qubit POVM \( \mathcal M_\ell
=
\{E_{\ell,a}\}_a \) at site \(\ell\). Define
\begin{equation}
x_\ell
:=
\frac{1}{2}
\sum_{a:\,\Tr(E_{\ell,a})>0}
\frac{
\Tr(E_{\ell,a}X)^2
}{
\Tr(E_{\ell,a})
},
\qquad
y_\ell
:=
\frac{1}{2}
\sum_{a:\,\Tr(E_{\ell,a})>0}
\frac{
\Tr(E_{\ell,a}Y)^2
}{
\Tr(E_{\ell,a})
}.
\label{eq:binary_xy_information}
\end{equation}
The quantities \(x_\ell\) and \(y_\ell\) measure how much local Fisher sensitivity this POVM can provide in the \(X\) and \(Y\) directions, respectively.

\begin{lemma}[Local \(X/Y\) information budget]
\label{lem:binary_local_xy_budget}
Every single qubit POVM satisfies
\begin{equation}
x_\ell+y_\ell\leq1.
\label{eq:binary_xy_budget}
\end{equation}
\end{lemma}

\begin{proof}
For each nonzero POVM effect, use the standard Bloch representation
\[
E_{\ell,a}
=
\gamma_{\ell,a}
\left(
\Id_{1} + \mathbf r_{\ell,a}\cdot\boldsymbol\sigma
\right),
\qquad
\gamma_{\ell,a}\geq0,
\qquad
\|\mathbf r_{\ell,a}\|_2\leq1,
\]
where
\(\boldsymbol\sigma=(X,Y,Z)\) is the vector of Pauli matrices and
\(\mathbf r_{\ell,a}=(r_{\ell,a,x},r_{\ell,a,y},r_{\ell,a,z})\)
is the Bloch vector associated with the effect.
The completeness condition \( \sum_aE_{\ell,a}=\Id_{1} \) implies \( \sum_a\gamma_{\ell,a}=1 \). Moreover,
\[
\Tr(E_{\ell,a})=2\gamma_{\ell,a},
\qquad
\Tr(E_{\ell,a}X)
=
2\gamma_{\ell,a}r_{\ell,a,x},
\qquad
\Tr(E_{\ell,a}Y)
=
2\gamma_{\ell,a}r_{\ell,a,y}.
\]
Substituting into \eqref{eq:binary_xy_information} gives
\[
x_\ell
=
\sum_a
\gamma_{\ell,a}r_{\ell,a,x}^2,
\qquad
y_\ell
=
\sum_a
\gamma_{\ell,a}r_{\ell,a,y}^2.
\]
Therefore
\[
x_\ell+y_\ell
=
\sum_a
\gamma_{\ell,a}
\left(
r_{\ell,a,x}^2+r_{\ell,a,y}^2
\right)
\leq
\sum_a
\gamma_{\ell,a}
\|\mathbf r_{\ell,a}\|_2^2
\leq
1.
\]
\end{proof}

\begin{lemma}[Fisher information factorization at \(c=0\)]
\label{lem:binary_product_factorization}
For every product POVM \(\mathcal M\) and every
\(S\in\mathcal S_{L,K}\),
\begin{equation}
I_{\mathcal M}(S;0)
=
\prod_{\ell\in S}x_\ell
\prod_{\ell\notin S}y_\ell.
\label{eq:binary_product_fisher_factorization}
\end{equation}
\end{lemma}

\begin{proof}
Using the notation \(A_{\ell,S}\) defined in Section~\ref{sec:binary_state_family}, so that
\(P_S=\bigotimes_{\ell=1}^L A_{\ell,S}\), multiplicativity of the trace over tensor products gives
\[
p_{\mathbf a}(c)
=
\Tr\!\left[
E_{\mathbf a}\sigma_{S,c}
\right]
=
2^{-L}
\left[
\prod_{\ell=1}^L
\Tr(E_{\ell,a_\ell})
+
c
\prod_{\ell=1}^L
\Tr(E_{\ell,a_\ell}A_{\ell,S})
\right].
\]
Writing
\[
e_{\ell,a}
:=
\Tr(E_{\ell,a}),
\qquad
q_{\ell,a}
:=
\Tr(E_{\ell,a}A_{\ell,S}),
\]
we have
\[
p_{\mathbf a}(0)
=
2^{-L}
\prod_{\ell=1}^L e_{\ell,a_\ell},
\qquad
\partial_c p_{\mathbf a}(0)
=
2^{-L}
\prod_{\ell=1}^L q_{\ell,a_\ell}.
\]
Hence
\[
I_{\mathcal M}(S;0)
=
\sum_{\mathbf a}
\frac{
[\partial_c p_{\mathbf a}(0)]^2
}{
p_{\mathbf a}(0)
}
=
2^{-L}
\sum_{\mathbf a}
\frac{
\prod_{\ell=1}^L q_{\ell,a_\ell}^2
}{
\prod_{\ell=1}^L e_{\ell,a_\ell}
}.
\]
Because the sum over the multi-index \(\mathbf a\) factors into independent sums over the individual sites,
\[
I_{\mathcal M}(S;0)
=
2^{-L}
\prod_{\ell=1}^L
\left[
\sum_{a_\ell}
\frac{
q_{\ell,a_\ell}^2
}{
e_{\ell,a_\ell}
}
\right].
\]
By \eqref{eq:binary_xy_information}, the bracket equals \(2x_\ell\) if
\(\ell\in S\) and \(2y_\ell\) if \(\ell\notin S\). Therefore
\[
I_{\mathcal M}(S;0)
=
2^{-L}
\prod_{\ell\in S}(2x_\ell)
\prod_{\ell\notin S}(2y_\ell)
=
\prod_{\ell\in S}x_\ell
\prod_{\ell\notin S}y_\ell,
\]
which proves \eqref{eq:binary_product_fisher_factorization}.
\end{proof}

\begin{lemma}[Support summed Fisher information budget]
\label{lem:binary_support_fisher_budget}
For every product POVM \(\mathcal M\),
\begin{equation}
\sum_{S\in\mathcal S_{L,K}}
I_{\mathcal M}(S;0)
\leq
1.
\label{eq:binary_support_sum}
\end{equation}
\end{lemma}

\begin{proof}
Let \( s_\ell:=x_\ell+y_\ell\leq1 \). If some \(s_\ell=0\), every term in the sum vanishes and the claim is trivial. Otherwise, set \( 
\beta_\ell
:=
\frac{x_\ell}{s_\ell}
\in[0,1] \). By Lemma~\ref{lem:binary_product_factorization},
\[
\sum_{S\in\mathcal S_{L,K}}
I_{\mathcal M}(S;0)
=
\left(
\prod_{\ell=1}^L s_\ell
\right)
\sum_{|S|=K}
\prod_{\ell\in S}\beta_\ell
\prod_{\ell\notin S}(1-\beta_\ell).
\]
The second factor is the probability that \( \sum_{\ell=1}^L B_\ell=K \), where \( B_\ell\sim\mathrm{Bernoulli}(\beta_\ell) \) independently. Therefore,
\[
\sum_{S\in\mathcal S_{L,K}}
I_{\mathcal M}(S;0)
=
\left(
\prod_{\ell=1}^L s_\ell
\right)
\Pr\left(
\sum_{\ell=1}^L B_\ell=K
\right)
\leq1,
\]
using
\[
\prod_{\ell=1}^L s_\ell\leq1
\qquad\text{and}\qquad
\Pr\left(
\sum_{\ell=1}^L B_\ell=K
\right)\leq1.
\]
\end{proof}

\begin{lemma}[Fisher information near \(c=0\)]
\label{lem:binary_product_nearzero}
For every product POVM \(\mathcal M\), every
\(S\in\mathcal S_{L,K}\), and every \(|c|<1\),
\begin{equation}
I_{\mathcal M}(S;c)
\leq
\frac{
I_{\mathcal M}(S;0)
}{
1-|c|
}.
\label{eq:binary_nearzero_bound}
\end{equation}
\end{lemma}

\begin{proof}
For each outcome, write \( p_{\mathbf a}(c)
=
p_{\mathbf a}(0)+c q_{\mathbf a} \), where \( q_{\mathbf a}
=
2^{-L}\Tr(E_{\mathbf a}P_S) \). Since \( -\Id_{L}\preceq P_S\preceq\Id_{L} \) and \(E_{\mathbf a}\succeq0\), we have \( |q_{\mathbf a}|
\leq
p_{\mathbf a}(0) \). If \(p_{\mathbf a}(0)=0\), then the preceding inequality implies
\(q_{\mathbf a}=0\), so this outcome has zero probability for every
\(c\) and contributes nothing to the Fisher information. We may
therefore restrict the sum below to outcomes with
\(p_{\mathbf a}(0)>0\).

For every such outcome, \( p_{\mathbf a}(c)
\geq
(1-|c|)p_{\mathbf a}(0) \). Because \( \partial_c p_{\mathbf a}(c)
=
q_{\mathbf a} \), we obtain
\[
I_{\mathcal M}(S;c)
=
\sum_{\mathbf a}
\frac{
q_{\mathbf a}^2
}{
p_{\mathbf a}(c)
}
\leq
\frac{1}{1-|c|}
\sum_{\mathbf a}
\frac{
q_{\mathbf a}^2
}{
p_{\mathbf a}(0)
}
=
\frac{
I_{\mathcal M}(S;0)
}{
1-|c|
}.
\]
\end{proof}

\paragraph{Theorem~\ref{thm:binary_product_povm_separation}}
\begin{proof}
The adaptive protocol described in
Section~\ref{sec:binary_localization} uses only product Pauli measurements and is therefore also an admissible product POVM protocol. Setting \(c_0=0\) in
\eqref{eq:binary_adaptive_constant} gives
\[
\limsup_{N\to\infty}
N
R_{\mathrm{AD,Prod}}^{(L,K)}(N;0)
\leq1.
\]
Conversely, the quantum Fisher information argument used in the proof of
Theorem~\ref{thm:binary_pauli_separation} applies to every single copy measurement, and therefore also to arbitrary adaptive product POVMs. It gives
\[
\liminf_{N\to\infty}
N
R_{\mathrm{AD,Prod}}^{(L,K)}(N;0)
\geq1,
\]
proving \eqref{eq:binary_product_adaptive}.

Now fix any nonadaptive product POVM protocol. For copy \(t\), let \( \mathcal M^{(t)} \) denote its signal POVM. By
Lemma~\ref{lem:binary_support_fisher_budget}, \( \sum_{S\in\mathcal S_{L,K}}
I_{\mathcal M^{(t)}}(S;0)
\leq1 \). Summing over all \(N\) copies gives \( 
\sum_{S\in\mathcal S_{L,K}}
J_S(0)
\leq
N \), where \( J_S(c)
=
\sum_{t=1}^N
I_{\mathcal M^{(t)}}(S;c) \). The same inequality holds after averaging over any pre-randomization of the measurement schedule.

Therefore, for every protocol there exists a support \( S_N^\star\in\mathcal S_{L,K} \) such that
\begin{equation}
J_{S_N^\star}(0)
\leq
\frac{N}{M_{L,K}}.
\label{eq:binary_product_worst_support}
\end{equation}

Marker outcomes can help the estimator infer \(S_N^\star\), but their distribution does not depend on \(c\), so they contribute no Fisher information for estimating \(c\). By
Lemma~\ref{lem:binary_product_nearzero}, for every
\(|c|\leq w_N\), we have \( J_{S_N^\star}(c)
\leq
\frac{
N
}{
M_{L,K}(1-w_N)
} \). Since \(w_N\to0\),
\[
\frac{1}{
M_{L,K}(1-w_N)
}
=
\frac{1}{M_{L,K}}
+
o(1).
\]
Hence, uniformly over \(|c|\leq w_N\),
\[
J_{S_N^\star}(c)
\leq
N
\left(
\frac{1}{M_{L,K}}
+
o(1)
\right).
\]
Applying Lemma~\ref{lem:local_fisher_to_mse} with \( A
=
\frac{1}{M_{L,K}} \) gives
\[
\liminf_{N\to\infty}
N
\sup_{|c|\leq w_N}
\E_{S_N^\star,c}
\left[
(\widehat c_N-c)^2
\right]
\geq
M_{L,K}.
\]
Since the local minimax criterion takes the worst case over \(S\), this bound holds for every nonadaptive product POVM protocol and estimator. Taking the infimum over them proves
\eqref{eq:binary_product_nonadaptive}.

The exponential scaling then follows from
\eqref{eq:binary_exponential_regime}.
\end{proof}

\subsection{Global single copy measurements}
\label{app:binary_global}

\paragraph{Theorem~\ref{thm:binary_global_no_separation}}
\begin{proof}
Consider the fixed nonadaptive protocol described above. For fixed \(S\) and \(c\), the product form \(\rho_{S,c}=\tau_S\otimes\sigma_{S,c}\), together with the product measurement across the marker and signal registers, implies that the joint outcome distribution factorizes. Hence the marker data used to construct \(\widehat S\) are independent of the recorded signal outcomes.

On the event \(\widehat S=S\), we have \(V_t=\xi_S(z_t)\). By \eqref{eq:binary_global_candidate_statistics}, \( \E_{S,c}[V_t]=c \). Since \(V_t\in\{\pm1\}\), we have \( \operatorname{Var}_{S,c}(V_t)
= 1-c^2 \). The \(N\) signal outcomes are independent across copies, so their sample mean is unbiased with MSE \( \frac{1-c^2}{N} \).

On the event \(\widehat S\neq S\), both the true parameter and the clipped estimate lie in \([-c_{\max},c_{\max}]\), so the squared error is at most \(4c_{\max}^2\). Therefore, uniformly over \(S\),
\[
\sup_{\substack{
c\in\Theta\\
|c-c_0|\leq w_N
}}
\E_{S,c}
\left[
(\widehat c_N-c)^2
\right]
\leq
\frac{
\sup_{|c-c_0|\leq w_N}(1-c^2)
}{N}
+
4c_{\max}^2
\Pr_S(\widehat S\neq S).
\]
Since \(w_N\to0\) and
\(N\Pr_S(\widehat S\neq S)\to0\),
\[
\limsup_{N\to\infty}
N
R_{\mathrm{NA,Glob}}^{(L,K)}(N;c_0)
\leq
1-c_0^2.
\]

For the converse, the quantum Fisher information argument used in the proof of Theorem~\ref{thm:binary_pauli_separation} does not depend on the measurement being product. For any fixed support \(S\), the marker state is independent of \(c\), while the signal state \(\sigma_{S,c}=2^{-L}(\Id_{L}+cP_S)\) has quantum Fisher information \(1/(1-c^2)\). Hence every single copy POVM, including an arbitrary global one, has classical Fisher information at most \(1/(1-c^2)\). For an adaptive protocol the conditional Fisher informations therefore accumulate to at most \( J_N(c) \leq \frac{N}{1-c^2} \). Applying Lemma~\ref{lem:local_fisher_to_mse} gives
\[
\liminf_{N\to\infty}
N
R_{\mathrm{AD,Glob}}^{(L,K)}(N;c_0)
\geq
1-c_0^2.
\]

Finally, every nonadaptive global protocol is also an admissible adaptive global protocol, so
\[
R_{\mathrm{AD,Glob}}^{(L,K)}(N;c_0)
\leq
R_{\mathrm{NA,Glob}}^{(L,K)}(N;c_0).
\]
Combining this ordering with the preceding upper and lower bounds proves \eqref{eq:binary_global_equal_constants}.
\end{proof}

\section{Proofs and Supporting Results for the Ternary Family}
\label{app:ternary_results}

\subsection{Adaptive benchmark}
\label{app:ternary_adaptive}

\paragraph{Lemma~\ref{lem:ternary_localization}}
\begin{proof}
At one marker site, the mean \(Z\)-outcome is \(-u\), \(0\), or \(+u\) when the local label is \(X\), \(Y\), or \(Z\), respectively. An incorrect decision therefore requires the sample mean to deviate from its true mean by at least \(u/2\). Hoeffding's inequality for \(\{\pm1\}\)-valued variables gives a per-site error probability at most \(2e^{-m_0u^2/8}\). A union bound over the \(L\) sites gives the first claim. Substituting the stated value of \(m_0\) gives the remaining conclusions.
\end{proof}

\paragraph{Theorem~\ref{thm:ternary_adaptive}}
\begin{proof}
On the event \(\widehat{\mathbf a}=\mathbf a\), each of the \(N-m_0\) parity samples has mean \(c\) and variance \(1-c^2\). Hence the clipped sample mean has conditional MSE at most \( \frac{1-c^2}{N-m_0}
\leq
\frac{1}{N-m_0} \). On the event \(\widehat{\mathbf a}\neq\mathbf a\), both the true parameter and the clipped estimate lie in \([-1,1]\), so the squared error is at most \(4\). Lemma~\ref{lem:ternary_localization} therefore gives, uniformly over \(\mathbf a\) and \(|c|\leq w_N\),
\[
\E_{\mathbf a,c}
\left[
(\widehat c_N-c)^2
\right]
\leq
\frac{1}{N-m_0}
+
4(N+2)^{-2}.
\]
Since \(m_0=o(N)\), multiplying by \(N\) and taking \(\limsup\) proves the result.
\end{proof}

\paragraph{Theorem~\ref{thm:ternary_adaptive_lower}}
\begin{proof}
Fix a hidden label \(\mathbf a\). The marker state \(\tau_{\mathbf a}\) is independent of \(c\), while the signal state \(\sigma_{\mathbf a,c}=2^{-L}(\Id_{L}+cP_{\mathbf a})\) has a \(c\)-independent eigenbasis and quantum Fisher information \( H(c)
=
\frac{1}{1-c^2} \). Thus every one copy measurement has classical Fisher information at most \(1/(1-c^2)\). For an adaptive protocol, let \(H_{t-1}\) denote the measurement history before copy \(t\). The Fisher information chain rule gives
\[
J_N(c)
=
\sum_{t=1}^N
\E\!\left[
I_t(c\mid H_{t-1})
\right]
\leq
\frac{N}{1-c^2}.
\]
Uniformly over \(|c|\leq w_N\), this is \(N(1+o(1))\). Lemma~\ref{lem:local_fisher_to_mse} therefore applies with \(A=1\). Since the local asymptotic minimax MSE takes the worst case over hidden labels, the resulting lower bound holds for every adaptive protocol and estimator, and taking the infimum over them proves the claim.
\end{proof}

\subsection{Aggregate Fisher information bounds}
\label{app:ternary_fisher_budget}

Throughout this subsection, \(d=2^L\) denotes the signal Hilbert space dimension.

\begin{lemma}[Pauli--Parseval identity]
\label{lem:pauli_parseval}
Let \(\{P_j\}_{j=0}^{d^2-1}\) be the \(L\)-qubit Pauli strings, with \(P_0=\Id_{L}\). For every Hermitian operator \(F\) on \((\mathbb C^2)^{\otimes L}\),
\[
\sum_{j=1}^{d^2-1}
\Tr(FP_j)^2
=
d\Tr(F^2)-\Tr(F)^2.
\]
\end{lemma}

\begin{proof}
The \(L\)-qubit Pauli strings form an orthogonal basis under the
Hilbert--Schmidt inner product. Indeed, for
\(\sigma,\sigma'\in\{\Id_{1},X,Y,Z\}\),
\(\Tr(\sigma\sigma')=2\delta_{\sigma\sigma'}\), and therefore, taking
\(L\)-fold tensor products, \( \Tr(P_jP_k)=d\,\delta_{jk}\), where \( d=2^L\). Since there are \(d^2\) Pauli strings, they form a basis for the space of
\(d\times d\) matrices. Thus any Hermitian \(F\) can be expanded as
\(F=\sum_j f_jP_j\). Taking the trace after multiplication by \(P_k\)
and using orthogonality gives \( f_k=\frac{\Tr(FP_k)}{d} \). Using the same orthogonality in the expansion of \(F^2\),
\[
\Tr(F^2)
=
d\sum_{j=0}^{d^2-1}f_j^2
=
\frac{1}{d}
\sum_{j=0}^{d^2-1}
\Tr(FP_j)^2.
\]
Finally, \(P_0=\Id_{L}\), so the \(j=0\) term is
\(\Tr(F)^2\). Multiplying by \(d\) and isolating this term yields \( \sum_{j=1}^{d^2-1}\Tr(FP_j)^2
=
d\Tr(F^2)-\Tr(F)^2 \), as claimed.
\end{proof}

\paragraph{Lemma~\ref{lem:ternary_signal_budget}}
\begin{proof}
At \(c=0\), \(\sigma_{\mathbf a,0}=\Id_{L}/d\) and \(\partial_c\sigma_{\mathbf a,c}|_{c=0}=P_{\mathbf a}/d\), so \( I_{\mathcal F}(\mathbf a;0)
=
\sum_y
\frac{\Tr(F_yP_{\mathbf a})^2}{d\Tr(F_y)} \), where any outcome with \(F_y=0\) is omitted from the sum. The same expression may be associated with every Pauli string \(P_j\), and all terms are nonnegative. Every ternary label gives a distinct non-identity Pauli string, so the sum over the \(3^L\) ternary labels is no larger than the sum over all \(d^2-1\) non-identity Pauli strings. Applying Lemma~\ref{lem:pauli_parseval} to each \(F_y\) gives
\[
\begin{aligned}
\sum_{j=1}^{d^2-1} I_{\mathcal F}(j;0)
&=
\sum_y
\left[
\frac{\Tr(F_y^2)}{\Tr(F_y)}
-
\frac{\Tr(F_y)}{d}
\right]\\
&\leq
\sum_y
\left[
\Tr(F_y)-\frac{\Tr(F_y)}{d}
\right]
=
d-1,
\end{aligned}
\]
where \(\Tr(F_y^2)\leq\Tr(F_y)^2\) for \(F_y\succeq0\), and \(\sum_y\Tr(F_y)=d\). Restricting the nonnegative sum to the ternary subset proves the result.
\end{proof}

\begin{lemma}[Marker side information factor]
\label{lem:ternary_marker_factor}
Let \(z\in\{\pm1\}^L\) index the marker \(Z\)-basis and define
\(
p(z\mid\mathbf a):=\langle z|\tau_{\mathbf a}|z\rangle
\). Then
\[
\sum_z
\max_{\mathbf a}
p(z\mid\mathbf a)
=
(1+u)^L.
\]
\end{lemma}

\begin{proof}
The maximization over \(\mathbf a\) factorizes across sites. From the definitions of \(\tau_X,\tau_Y,\tau_Z\),
\[
p(+1\mid X)=\frac{1-u}{2},
\qquad
p(+1\mid Y)=\frac12,
\qquad
p(+1\mid Z)=\frac{1+u}{2},
\]
with the roles of \(X\) and \(Z\) reversed for outcome \(-1\). Hence \(\max_a p(\pm1\mid a)=(1+u)/2\). Summing over \(z\) therefore factorizes into \(L\) identical single-site sums, each equal to \(1+u\), which proves the claim.
\end{proof}

\paragraph{Theorem~\ref{thm:ternary_joint_budget}}
\begin{proof}
Because \(\tau_{\mathbf a}\) is diagonal in the marker \(Z\)-basis, we have \( \rho_{\mathbf a,c}
=
\sum_z
p(z\mid\mathbf a)
|z\rangle\langle z|
\otimes
\sigma_{\mathbf a,c} \). For each outcome \(y\) and marker string \(z\), define the signal operator \( \widetilde E_{y,z}
:=
(\langle z|\otimes\Id_{L})
E_y
(|z\rangle\otimes\Id_{L}) \). Each \(\widetilde E_{y,z}\succeq0\), and for fixed \(z\), \(\sum_y\widetilde E_{y,z}=\Id_{L}\), so \(\mathcal M_z:=\{\widetilde E_{y,z}\}_y\) is a valid signal POVM. Moreover,
\[
p_y(\mathbf a,c)
=
\Tr(E_y\rho_{\mathbf a,c})
=
\sum_z
p(z\mid\mathbf a)
\Tr(\widetilde E_{y,z}\sigma_{\mathbf a,c}).
\]
Consider the refined experiment with outcomes \((z,y)\), whose probabilities are
\(
q_{z,y}(\mathbf a,c)
=
p(z\mid\mathbf a)\Tr(\widetilde E_{y,z}\sigma_{\mathbf a,c})
\). Its \(y\)-marginal is the original outcome distribution. For each \(y\), Cauchy--Schwarz gives
\[
\frac{
\left[
\partial_c\sum_z q_{z,y}(\mathbf a,c)
\right]^2
}{
\sum_z q_{z,y}(\mathbf a,c)
}
\leq
\sum_z
\frac{
\left[
\partial_c q_{z,y}(\mathbf a,c)
\right]^2
}{
q_{z,y}(\mathbf a,c)
},
\]
with zero-probability terms omitted as in \eqref{eq:generic_fisher_information}. Thus forgetting \(z\) cannot increase Fisher information. Since \(p(z\mid\mathbf a)\) is independent of \(c\), we have \( I_{\mathcal M}(\mathbf a;0)
\leq
\sum_z
p(z\mid\mathbf a)
I_{\mathcal M_z}(\mathbf a;0) \). Summing over \(\mathbf a\) and applying Lemmas~\ref{lem:ternary_signal_budget} and \ref{lem:ternary_marker_factor},
\[
\begin{aligned}
\sum_{\mathbf a}
I_{\mathcal M}(\mathbf a;0)
&\leq
\sum_z
\sum_{\mathbf a}
p(z\mid\mathbf a)
I_{\mathcal M_z}(\mathbf a;0)\\
&\leq
\sum_z
\left(
\max_{\mathbf b}p(z\mid\mathbf b)
\right)
\sum_{\mathbf a}
I_{\mathcal M_z}(\mathbf a;0)\\
&\leq
(2^L-1)
\sum_z
\max_{\mathbf b}p(z\mid\mathbf b) = (2^L-1)(1+u)^L.
\end{aligned}
\]
\end{proof}

\begin{corollary}[Exponentially small worst label information]
\label{cor:ternary_worst_word}
For every global single copy POVM, there exists a hidden label \(\mathbf a^\star\in\{X,Y,Z\}^L\) such that
\[
I_{\mathcal M}(\mathbf a^\star;0)
\leq
B_L.
\]
\end{corollary}

\begin{proof}
Average \eqref{eq:ternary_joint_budget} over the \(3^L\) hidden labels and use \eqref{eq:ternary_BL}.
\end{proof}

\subsection{Local window control and the nonadaptive lower bound}
\label{app:ternary_nonadaptive}

\begin{lemma}[Near-zero Fisher information bound]
\label{lem:ternary_nearzero}
For every global single copy POVM \(\mathcal M=\{E_y\}_y\), every hidden label \(\mathbf a\), and every \(|c|<1\),
\[
I_{\mathcal M}(\mathbf a;c)
\leq
\frac{I_{\mathcal M}(\mathbf a;0)}{1-|c|}.
\]
\end{lemma}

\begin{proof}
Write
\[
\rho_{\mathbf a,c}
=
\rho_{\mathbf a,0}
+c\dot\rho_{\mathbf a},
\qquad
\rho_{\mathbf a,0}
=
\tau_{\mathbf a}\otimes\frac{\Id_{L}}{d},
\qquad
\dot\rho_{\mathbf a}
=
\tau_{\mathbf a}\otimes\frac{P_{\mathbf a}}{d}.
\]
Since \(-\Id_{L}\preceq P_{\mathbf a}\preceq\Id_{L}\), we have \( -\rho_{\mathbf a,0}
\preceq
\dot\rho_{\mathbf a}
\preceq
\rho_{\mathbf a,0} \). For \(E_y\succeq0\), set
\(
p_{y,0}=\Tr(E_y\rho_{\mathbf a,0})
\) and
\(
q_y=\Tr(E_y\dot\rho_{\mathbf a})
\). The operator inequality gives \(|q_y|\leq p_{y,0}\). If \(p_{y,0}=0\), then \(q_y=0\) and that outcome contributes nothing. Otherwise, \( p_y(c)
=
p_{y,0}+cq_y
\geq
(1-|c|)p_{y,0} \), and therefore
\[
I_{\mathcal M}(\mathbf a;c)
=
\sum_y
\frac{q_y^2}{p_y(c)}
\leq
\frac{1}{1-|c|}
\sum_y
\frac{q_y^2}{p_{y,0}}
=
\frac{I_{\mathcal M}(\mathbf a;0)}{1-|c|}.
\]
\end{proof}

For a deterministic nonadaptive sequence \(\mathcal M^{(1)},\ldots,\mathcal M^{(N)}\) of global single copy POVMs, define \( J_{\mathbf a}(c)
:=
\sum_{t=1}^N
I_{\mathcal M^{(t)}}(\mathbf a;c) \). 

\begin{lemma}[Worst hidden label for an \(N\)-copy nonadaptive protocol]
\label{lem:ternary_worst_word_N}
For every deterministic nonadaptive sequence, there exists \(\mathbf a_N^\star\in\{X,Y,Z\}^L\) such that \( J_{\mathbf a_N^\star}(0)
\leq
N B_L \). If the measurement schedule is instead randomized in advance using a random variable \(R\), independent of the unknown state and recorded as part of the experimental data, the same conclusion holds with
\(
\overline J_{\mathbf a}(0)
:=
\E_R[J_{\mathbf a}^{(R)}(0)]
\)
in place of \(J_{\mathbf a}(0)\).
\end{lemma}

\begin{proof}
For a deterministic sequence, Theorem~\ref{thm:ternary_joint_budget} gives
\[
\sum_{\mathbf a}
J_{\mathbf a}(0)
=
\sum_{t=1}^N
\sum_{\mathbf a}
I_{\mathcal M^{(t)}}(\mathbf a;0)
\leq
N(2^L-1)(1+u)^L
=
N\,3^L B_L.
\]
Averaging over the \(3^L\) labels proves the first claim. For a randomized schedule, take expectation over \(R\) before summing over \(\mathbf a\); since \(R\) is independent of the state, the same bound applies to \(\overline J_{\mathbf a}(0)\).
\end{proof}

\paragraph{Theorem~\ref{thm:ternary_nonadaptive}}
\begin{proof}
Fix an arbitrary nonadaptive protocol and estimator. If the measurement schedule is randomized in advance, include the realized schedule \(R\) in the observation record. Because \(R\) is independent of \(c\), the total Fisher information of the recorded experiment is the average of the conditional Fisher informations over \(R\).

Let \(\mathbf a_N^\star\) be the label supplied by Lemma~\ref{lem:ternary_worst_word_N}. Applying Lemma~\ref{lem:ternary_nearzero} to each one-copy measurement and, when needed, averaging over \(R\), gives, uniformly for \(|c|\leq w_N\), \( J_{\mathbf a_N^\star}(c)
\leq
\frac{N B_L}{1-w_N}
=
N\bigl(B_L+o(1)\bigr) \). Here \(J\) is understood as \(\overline J\) in the randomized case, and the \(o(1)\) term depends only on \(w_N\), so it is uniform over admissible nonadaptive protocols. Lemma~\ref{lem:local_fisher_to_mse} therefore applies with \(A=B_L\), yielding
\[
\liminf_{N\to\infty}
N
\sup_{|c|\leq w_N}
\E_{\mathbf a_N^\star,c}
\left[
(\widehat c_N-c)^2
\right]
\geq
\frac{1}{B_L}.
\]
The hidden label may depend on the fixed protocol and on \(N\), but it is selected before any data are observed. Since the local asymptotic minimax MSE takes the worst case over hidden labels after the protocol and estimator have been chosen, this lower bound holds for every nonadaptive protocol and estimator. Taking the infimum over them proves the theorem.
\end{proof}

\subsection{Finite sample separation and the 2-design benchmark}
\label{app:ternary_finite_sample}

\paragraph{Theorem~\ref{thm:ternary_finite_sample}}
\begin{proof}
\emph{Adaptive upper bound.} Use \(m_0=\lceil 8u^{-2}\log(12L)\rceil\) copies for marker localization; by Lemma~\ref{lem:ternary_localization}, the label is wrong with probability at most \(2L\exp(-m_0u^2/8)\leq1/6\), uniformly in \(\mathbf a\) and \(c\). On \(m_1=\lceil2\epsilon^{-2}\log12\rceil\) further copies, measure each signal qubit in the Pauli basis selected by the estimated label and let \(\widehat c\) be the mean of the parity outcomes. Conditional on correct localization these are independent \(\{\pm1\}\)-valued variables with mean \(c\), so Hoeffding's inequality gives \(\Pr(|\widehat c-c|>\epsilon\mid\widehat{\mathbf a}=\mathbf a)\leq2\exp(-m_1\epsilon^2/2)\leq1/6\). A union bound over the two failure events proves \eqref{eq:ternary_finite_adaptive}.

\emph{Nonadaptive lower bound.} Fix a label \(\mathbf a\) and one global single copy POVM \(\mathcal M=\{E_y\}\). As in the proof of Lemma~\ref{lem:ternary_nearzero}, \(p_y(c)=\alpha_y+c\beta_y\) with \(|\beta_y|\leq\alpha_y\) and \(I_{\mathcal M}(\mathbf a;0)=\sum_{y:\alpha_y>0}\beta_y^2/\alpha_y\). For \(0<t<1\), \(p_y(-t)\geq(1-t)\alpha_y\), and the chi-square bound on relative entropy gives
\begin{equation}
D_{\mathrm{KL}}\bigl(p(t)\,\|\,p(-t)\bigr)
\leq\sum_{y:\alpha_y>0}\frac{\bigl(p_y(t)-p_y(-t)\bigr)^2}{p_y(-t)}
\leq\frac{4t^2}{1-t}\sum_{y:\alpha_y>0}\frac{\beta_y^2}{\alpha_y}
=\frac{4t^2}{1-t}\,I_{\mathcal M}(\mathbf a;0);
\label{eq:ternary_two_point_KL}
\end{equation}
outcomes with \(\alpha_y=0\) have zero probability at both parameter values and may be dropped. Now fix a nonadaptive protocol with \(N\) copies and include its state independent seed \(R\) in the record. By Lemma~\ref{lem:ternary_worst_word_N} there is a label \(\mathbf a^\star\), chosen before any data are observed, with \(\E_R[\sum_{i=1}^NI_{\mathcal M^{(i)}(R)}(\mathbf a^\star;0)]\leq NB_L\). Conditional on \(R\) the outcomes on different copies are independent, so relative entropies add, and \(R\) has the same law at both parameter values; hence the recorded transcript laws \(P_t\) and \(P_{-t}\) at the label \(\mathbf a^\star\) satisfy
\begin{equation}
D_{\mathrm{KL}}(P_t\,\|\,P_{-t})\leq\frac{4t^2}{1-t}\,NB_L .
\label{eq:ternary_transcript_KL}
\end{equation}
Randomness used by the estimator can be included in \(R\) as well. Take \(t=2\epsilon\), so that \(\pm t\in\Theta\) and \(t\leq\tfrac12\). An estimator satisfying \eqref{eq:ternary_finite_accuracy} yields a test between \(c=t\) and \(c=-t\), deciding \(c=t\) when \(\widehat c>0\). Whenever \(|\widehat c-c|\leq\epsilon\) this decision is correct, so both error probabilities are at most \(\tfrac13\). Hence \(\mathrm{TV}(P_t,P_{-t})\geq\tfrac13\), and Pinsker's inequality gives \(D_{\mathrm{KL}}(P_t\,\|\,P_{-t})\geq2\,\mathrm{TV}(P_t,P_{-t})^2\geq\tfrac29\). Since \(4t^2/(1-t)\leq32\epsilon^2\), \eqref{eq:ternary_transcript_KL} gives \(\tfrac29\leq32\epsilon^2NB_L\), which is \eqref{eq:ternary_finite_nonadaptive}.
\end{proof}

\paragraph{Proposition~\ref{prop:ternary_nonadaptive_upper}}
\begin{proof}
Write \(d=2^L\) and \(g_y(\mathbf a):=\langle\psi_y|P_{\mathbf a}|\psi_y\rangle\in[-1,1]\). Then \(p_y(\mathbf a,c)=\Tr(F_y\sigma_{\mathbf a,c})=|\mathcal Y|^{-1}\bigl(1+c\,g_y(\mathbf a)\bigr)\), so
\[
I_{\mathcal F}(\mathbf a;c)
=
\frac{1}{|\mathcal Y|}\sum_y\frac{g_y(\mathbf a)^2}{1+c\,g_y(\mathbf a)}
\geq
\frac{1}{1+|c|}\cdot\frac{1}{|\mathcal Y|}\sum_yg_y(\mathbf a)^2
=
\frac{1}{(d+1)(1+|c|)},
\]
by \eqref{eq:two_design_moments} with \(A=P_{\mathbf a}\) and \(\Tr(P_{\mathbf a}^2)=d\); equality holds at \(c=0\).

For the estimator, first form \(\widehat{\mathbf a}\) from all \(N\) marker outcomes as in Lemma~\ref{lem:ternary_localization} with \(m_0=N\), so that \(\Pr_{\mathbf a}(\widehat{\mathbf a}\neq\mathbf a)\leq2L\exp(-Nu^2/8)=o(N^{-1})\). Then set \(T_t:=(d+1)\,g_{y_t}(\widehat{\mathbf a})\), where \(y_t\) is the signal outcome on copy \(t\), and let \(\widehat c_N\) be the sample mean of \(T_1,\ldots,T_N\) clipped to \([-1,1]\). The marker and signal outcomes are independent, because both the state and the measurement are products across the two registers. On the event \(\widehat{\mathbf a}=\mathbf a\), \eqref{eq:two_design_moments} gives \(\E_{\mathbf a,c}[T_t\mid\widehat{\mathbf a}=\mathbf a]=(d+1)\bigl(0+c/(d+1)\bigr)=c\) and, using \(|g_y|\leq1\),
\[
\operatorname{Var}_{\mathbf a,c}(T_t\mid\widehat{\mathbf a}=\mathbf a)
\leq
(d+1)^2\,\frac{1}{|\mathcal Y|}\sum_yg_y(\mathbf a)^2\bigl(1+c\,g_y(\mathbf a)\bigr)
\leq
(d+1)^2\,\frac{1+|c|}{d+1}
=
(d+1)(1+|c|).
\]
Hence, uniformly over \(|c|\leq w_N\), the conditional MSE on this event is at most \((d+1)(1+w_N)/N\), and clipping cannot increase it. On the complementary event the squared error is at most \(4\). Therefore, uniformly over \(\mathbf a\),
\[
\sup_{|c|\leq w_N}\E_{\mathbf a,c}\bigl[(\widehat c_N-c)^2\bigr]
\leq
\frac{(d+1)(1+w_N)}{N}+8L\,e^{-Nu^2/8}.
\]
Multiplying by \(N\) and letting \(N\to\infty\) proves the claim. The final chain of inequalities combines this with Theorem~\ref{thm:ternary_nonadaptive}.
\end{proof}

\section{Proofs of the Structural Results}
\label{app:structural_results}

\subsection{Classicality and signal correlations}
\label{app:classicality}

\paragraph{Lemma~\ref{lem:full_separability}}
\begin{proof}
For each site \(\ell\), choose an orthonormal eigenbasis
\(
\{|s_\ell\rangle_{w_\ell}:s_\ell\in\{\pm1\}\}
\)
of \(w_\ell\). Since \(w_\ell\in\{X,Y,Z\}\), its eigenvalues are
\(\pm1\), so the basis vectors may be labeled such that \( w_\ell |s_\ell\rangle_{w_\ell}
=
s_\ell |s_\ell\rangle_{w_\ell} \). For
\(
\mathbf s=(s_1,\ldots,s_L)\in\{\pm1\}^L
\),
define the product state \( |\mathbf s\rangle_{\mathbf w}
=
\bigotimes_{\ell=1}^L
|s_\ell\rangle_{w_\ell} \). Using
\(
P_{\mathbf w}
=
\bigotimes_{\ell=1}^L w_\ell
\),
the action of \(P_{\mathbf w}\) on this product state factorizes:
\[
P_{\mathbf w}|\mathbf s\rangle_{\mathbf w}
=
\bigotimes_{\ell=1}^L
\left(
w_\ell|s_\ell\rangle_{w_\ell}
\right) =
\bigotimes_{\ell=1}^L
\left(
s_\ell|s_\ell\rangle_{w_\ell}
\right) =
\left(
\prod_{\ell=1}^L s_\ell
\right)
|\mathbf s\rangle_{\mathbf w}.
\]
Thus the product states
\(
\{|\mathbf s\rangle_{\mathbf w}\}_{\mathbf s}
\)
form an eigenbasis of \(P_{\mathbf w}\), with eigenvalues
\(
\prod_{\ell=1}^L s_\ell
\).
Consequently, \( \Id_{L}
=
\sum_{\mathbf s\in\{\pm1\}^L}
|\mathbf s\rangle_{\mathbf w}
\langle\mathbf s|_{\mathbf w} \), and \( P_{\mathbf w}
=
\sum_{\mathbf s\in\{\pm1\}^L}
\left(
\prod_{\ell=1}^L s_\ell
\right)
|\mathbf s\rangle_{\mathbf w}
\langle\mathbf s|_{\mathbf w} \). Substituting these two decompositions into
\(
\sigma_{\mathbf w,c}
=
2^{-L}(\Id_{L}+cP_{\mathbf w})
\)
gives
\begin{equation}
\sigma_{\mathbf w,c}
=
\sum_{\mathbf s\in\{\pm1\}^L}
\frac{
1+c\prod_{\ell=1}^L s_\ell
}{
2^L
}
|\mathbf s\rangle_{\mathbf w}
\langle\mathbf s|_{\mathbf w}.
\label{eq:separable_signal_decomposition}
\end{equation}

For every \(\mathbf s\),
\(
\prod_{\ell=1}^L s_\ell\in\{\pm1\}
\),
so the coefficients in
\eqref{eq:separable_signal_decomposition} are nonnegative whenever
\(|c|\leq1\). They also sum to one, since
\[
\sum_{\mathbf s}
\frac{
1+c\prod_{\ell=1}^L s_\ell
}{
2^L
}
=
\frac{1}{2^L}
\left[
2^L
+
c
\sum_{\mathbf s}
\prod_{\ell=1}^L s_\ell
\right] =
\frac{1}{2^L}
\left[
2^L
+
c
\prod_{\ell=1}^L
\left(
\sum_{s_\ell=\pm1}s_\ell
\right)
\right] =
1.
\]
Finally, each projector in
\eqref{eq:separable_signal_decomposition} is itself a product state, \( |\mathbf s\rangle_{\mathbf w}
\langle\mathbf s|_{\mathbf w}
=
\bigotimes_{\ell=1}^L
|s_\ell\rangle_{w_\ell}
\langle s_\ell|_{w_\ell} \). Therefore
\(\sigma_{\mathbf w,c}\) is a convex combination of product states and is fully separable across the \(L\) signal qubits. Moreover, the projectors \(|\mathbf s\rangle_{\mathbf w}\langle\mathbf s|_{\mathbf w}\) are mutually orthogonal, so \eqref{eq:separable_signal_decomposition} is the spectral decomposition of \(\sigma_{\mathbf w,c}\) in a product of local orthonormal bases: \(\sigma_{\mathbf w,c}\) is a classical probability distribution on \(\{\pm1\}^L\) written in that basis. Measuring any subset of the qubits in this basis leaves the state unchanged, so the classical correlations equal the mutual information and the quantum discord vanishes across every bipartition and in the multipartite sense~\cite{OllivierZurek2001,HendersonVedral2001,RulliSarandy2011,ModiBrodutchCablePaterekVedral2012}; no further argument is needed, since this is the defining property of such states.
\end{proof}

\paragraph{Corollary~\ref{cor:family_separability}}
\begin{proof}
For the binary family, take
\(
w_\ell=X
\)
when \(\ell\in S\) and
\(
w_\ell=Y
\)
otherwise. Lemma~\ref{lem:full_separability} shows that
\(\sigma_{S,c}\) is fully separable, while
\(\tau_S\) in \eqref{eq:binary_marker_state} is already a product state across the marker qubits. Thus
\(
\rho_{S,c}=\tau_S\otimes\sigma_{S,c}
\)
is fully separable across all \(2L\) qubits.

For the ternary family, take
\(
w_\ell=a_\ell
\).
Again Lemma~\ref{lem:full_separability} gives full separability of
\(\sigma_{\mathbf a,c}\), and
\(
\tau_{\mathbf a}
=
\bigotimes_{\ell=1}^L\tau_{a_\ell}
\)
is a product marker state. Hence
\(
\rho_{\mathbf a,c}
=
\tau_{\mathbf a}\otimes\sigma_{\mathbf a,c}
\)
is fully separable as well. In both cases the marker state is diagonal in the \(Z\) basis, so the product \(\tau\otimes\sigma\) is diagonal in the stated local product basis and is classical by the argument of Lemma~\ref{lem:full_separability}.
\end{proof}

\paragraph{Proposition~\ref{prop:signal_correlation_structure}}
\begin{proof}
Since every nonidentity Pauli string is traceless and
\(P_{\mathbf w}^2=\Id_{L}\), \[
\Tr\!\left(P_{\mathbf w}\sigma_{\mathbf w,c}\right)
=
\frac{1}{2^L}
\left[
\Tr(P_{\mathbf w})
+
c\Tr(P_{\mathbf w}^2)
\right]
=
c,
\]
which proves \eqref{eq:signal_full_correlation}.

Now let
\(T\subsetneq\{1,\ldots,L\}\).
Because \(T^c\) is nonempty, the partial trace of
\(P_{\mathbf w}\) over \(T^c\) vanishes: at least one Pauli factor is traced out, and every Pauli matrix \(X,Y,Z\) is traceless. Therefore
\[
\Tr_{T^c}\!\left(\sigma_{\mathbf w,c}\right)
=
\frac{1}{2^L}
\Tr_{T^c}(\Id_{L})
=
\frac{\Id_{|T|}}{2^{|T|}},
\]
which proves \eqref{eq:signal_proper_marginal}.

Finally, when \(L\geq2\), every single qubit marginal is \(\Id_{1}/2\). If
\(\sigma_{\mathbf w,c}\) were a product state across the signal qubits, it would therefore equal the product of these marginals,
\(
\Id_{L}/2^L
\).
For \(c\neq0\), however,
\(
\sigma_{\mathbf w,c}
=
2^{-L}(\Id_{L}+cP_{\mathbf w})
\neq
\Id_{L}/2^L
\).
Hence the state is not a product state.
\end{proof}

\paragraph{Proposition~\ref{prop:commuting_simulation}}
\begin{proof}
Simultaneous diagonalization gives an orthonormal basis \(\{|x\rangle\}_x\), independent of \(\theta\), such that \(\omega_\theta=\sum_xp_\theta(x)|x\rangle\langle x|\). The nonadaptive protocol measures every copy in this basis, obtaining independent samples \(X_1,\ldots,X_N\sim p_\theta\), and then simulates the adaptive transcript in post processing. If, after a simulated history \(h\), the adaptive protocol would apply the POVM \(\{E_{h,y}\}_y\) to copy \(t\), draw the next simulated outcome from \(K_h(y\mid X_t)=\langle X_t|E_{h,y}|X_t\rangle\). This is a valid stochastic kernel because \(E_{h,y}\succeq0\) and \(\sum_yE_{h,y}=\Id\). Since \(X_t\) is independent of the simulated past, \(\Pr(Y_t=y\mid h)=\sum_xp_\theta(x)\langle x|E_{h,y}|x\rangle=\Tr(E_{h,y}\omega_\theta)\), which is the conditional outcome law of the adaptive protocol. By induction the two transcripts have the same joint distribution. Any internal randomness of the adaptive protocol is generated as part of the simulation, and the same estimator is applied. Thus every adaptive protocol has a nonadaptive simulation with the same risk. Since every nonadaptive protocol is also adaptive, the optimal risks coincide.
\end{proof}

\paragraph{Proposition~\ref{prop:classical_restricted}}
\begin{proof}
If \(\mathbf r=\mathbf a\), the observed cell bits are the live bits. Their parity is a \(\{\pm1\}\)-valued variable with mean \(c\), whose Fisher information is \(1/(1-c^2)\). If \(r_\ell\neq a_\ell\) at some site, the observed bit there is an independent uniform decoy, while the observed live bits at the other sites form a proper subset whose marginal law is uniform because only the full parity is biased. The cell record is then independent of \(c\). The marker bits are also independent of \(c\).

A fully observed sample has Fisher information \(1/(1-c^2)\) about \(c\). By data processing and the Fisher information chain rule, any adaptive protocol using \(N\) samples has total Fisher information at most \(N/(1-c^2)\). Uniformly over the local window around \(c_0\), Lemma~\ref{lem:local_fisher_to_mse} therefore gives \( \liminf_{N\to\infty}N R_{\mathrm{AD}}(N;c_0)\geq 1-c_0^2 \) for any allowed protocol.

For the matching adaptive upper bound, use \(m_0=O(\log N)\) samples to identify each \(a_\ell\) from the sign of its marker sample mean. Hoeffding's inequality and a union bound give \(\Pr(\widehat{\mathbf a}\neq\mathbf a)=o(N^{-1})\) for a suitable constant in \(m_0\). On the remaining samples, use pattern \(\widehat{\mathbf a}\) and estimate \(c\) from the parity sample mean. The same clipped-estimator argument as in Theorem~\ref{thm:ternary_adaptive} gives \( \limsup_{N\to\infty} N R_{\mathrm{AD}}(N;c_0) \leq 1-c_0^2 \). Together with the preceding lower bound, this proves \( \lim_{N\to\infty} N R_{\mathrm{AD}}(N;c_0) = 1-c_0^2 \).

For a nonadaptive protocol, let \(n_{\mathbf b}\) be the expected number of samples read with pattern \(\mathbf b\), including any randomization fixed in advance. Since \( \sum_{\mathbf b}n_{\mathbf b}=N \), there is a label \(\mathbf a^\star\) with \(n_{\mathbf a^\star}\leq N2^{-L} \). Conditioning on the recorded schedule, only samples measured with pattern \(\mathbf a^\star\) carry Fisher information about \(c\). Because the schedule is independent of \(c\), averaging over it gives total Fisher information at most \( \frac{n_{\mathbf a^\star}}{1-c^2} \leq \frac{N2^{-L}}{1-c^2} = N\left( \frac{2^{-L}}{1-c_0^2}+o(1) \right) \) uniformly over \(|c-c_0|\leq w_N\). Lemma~\ref{lem:local_fisher_to_mse} then yields \( \liminf_{N\to\infty} N R_{\mathrm{NA}}(N;c_0) \geq 2^L(1-c_0^2) \).

For the three-component extension, the marker means \(-u,0,+u\) allow the three local labels to be localized with exponentially small error in \(m_0\) by the same concentration argument as in Lemma~\ref{lem:ternary_localization}. The nonadaptive counting step then has \(3^L\) possible patterns, replacing the factor \(2^L\) by \(3^L\).
\end{proof}

\subsection{MPO representation}
\label{app:mpo}

\paragraph{Proposition~\ref{prop:ternary_mpo}}
\begin{proof}
Equation~\eqref{eq:ternary_two_product_mpo} already expresses the state as a sum of two tensor products of local block operators. For \(L\geq2\), an explicit MPO with open boundaries and bond dimension two is obtained from
\[
W^{[1]}
=
\begin{bmatrix}
A_1 & cB_1
\end{bmatrix},
\qquad
W^{[\ell]}
=
\begin{bmatrix}
A_\ell & 0\\
0 & B_\ell
\end{bmatrix},
\quad
2\leq\ell\leq L-1,
\quad \text{and} \quad
W^{[L]}
=
\begin{bmatrix}
A_L\\
B_L
\end{bmatrix}.
\]
Because the interior matrices are diagonal, their multiplication leaves
only two nonzero contributions. Selecting the upper diagonal entry at
every interior site gives \( A_1\otimes A_2\otimes\cdots\otimes A_L \), whereas selecting the lower diagonal entry at every interior site gives \( c\,B_1\otimes B_2\otimes\cdots\otimes B_L \). Any mixed selection necessarily uses an off-diagonal entry of one of the
interior matrices and therefore vanishes. Hence the contraction of the
\(W^{[\ell]}\) matrices gives \( \bigotimes_{\ell=1}^L A_\ell
+
c\bigotimes_{\ell=1}^L B_\ell \). Multiplying by the overall factor \(2^{-L}\) reproduces
\eqref{eq:ternary_two_product_mpo}. For \(L=1\), the representation is
immediate. Hence an exact MPO representation with bond dimension at
most two exists for every \(L\).
\end{proof}

\section{Proof of the Metric Relations}
\label{app:metric_relations}

\paragraph{Proposition~\ref{prop:ternary_metric_geometry}}
\begin{proof}
Writing \(\delta=c-c'\), and \( \rho_{\mathbf a,c}-\rho_{\mathbf a,c'}
=
\delta\,\tau_{\mathbf a}\otimes\frac{P_{\mathbf a}}{d} \). Since \(\tau_{\mathbf a}\) is a density matrix, \(\|\tau_{\mathbf a}\|_1=1\). Also \(P_{\mathbf a}\) is Hermitian and unitary on the \(d\)-dimensional signal space, so \(\|P_{\mathbf a}/d\|_1=1\). Multiplicativity of the trace norm under tensor products gives \eqref{eq:ternary_trace_metric_relation}.

For the Frobenius norm, the single qubit marker states satisfy
\[
\Tr(\tau_X^2)=\Tr(\tau_Z^2)=\frac{1+u^2}{2},
\qquad
\Tr(\tau_Y^2)=\frac12.
\]
Therefore \( \Tr(\tau_{\mathbf a}^2)
=
\frac{(1+u^2)^{m(\mathbf a)}}{d} \). Since \(P_{\mathbf a}^2=\Id_{L}\), we also have \( \left\|\frac{P_{\mathbf a}}{d}\right\|_F^2
=
\frac{1}{d} \). Multiplicativity of the Frobenius norm then gives \eqref{eq:ternary_frob_metric_relation}. Finally, \( \Tr\!\left[\left(\frac{\Id_{L}+cP_{\mathbf a}}{d}\right)^2\right]
=
\frac{1+c^2}{d} \), because \(\Tr(P_{\mathbf a})=0\). Combining this identity with \(\Tr(\tau_{\mathbf a}^2)\) yields \eqref{eq:ternary_state_frob_norm}, and \eqref{eq:ternary_relative_frob_relation} follows immediately.
\end{proof}

\section*{Acknowledgment}

The authors acknowledge the use of ChatGPT, Claude, and Gemini for analytical and grammatical feedback, as well as collaborative text generation. The scientific results were developed by the authors, who independently evaluated any suggestions before incorporating them.

\bibliographystyle{IEEEtran}
\bibliography{references}

\end{document}